\documentclass[12pt,final,journal,letterpaper,onecolumn]{IEEEtran}

\usepackage{amsmath}
\usepackage{amssymb}
\usepackage{amsfonts}
\usepackage{amsxtra}                    % Use various AMS packages
\usepackage{bm}
\usepackage{epsfig}
\usepackage{graphicx}
\usepackage{makeidx}  % allows for indexgeneration
\usepackage{paralist}
\usepackage{setspace}
\usepackage[active]{srcltx}
\usepackage{tensor}
\usepackage{url}
\usepackage{verbatim}
    \DeclareMathOperator*{\argmax}{arg\,max}

    \newtheorem{theorem}{Theorem}  
    \newtheorem{Cl}{Claim}
    \newtheorem{Ass}{Assumption}

    \newcommand{\bs}{\ensuremath{\backslash}}

    \newcommand{\dg}[1]{\ensuremath{\mb{diag}(#1)}}
    \newcommand{\tx}{\ensuremath{\;\exists\;}}
    \newcommand{\fn}[1]{\ensuremath{\footnote{#1}}}
    \newcommand{\fa}{\ensuremath{\;\forall\;}}
    \newcommand{\h}[1]{\ensuremath{\hat{#1}}}
    \newcommand{\hs}[1]{\ensuremath{\hspace{#1}}}

    \newcommand{\mb}[1]{\ensuremath{\mbox{#1}}}
    \newcommand{\D}{\ensuremath{\nabla}}              
    
    \newcommand{\nnb}{\ensuremath{\nonumber}}
    \newcommand{\nin}{\ensuremath{\notin}}
    
    \newcommand{\ol}[1]{\ensuremath{\overline{#1}}}
    \newcommand{\raro}{\ensuremath{\rightarrow}}
    \newcommand{\Raro}{\ensuremath{\Rightarrow}}

    \newcommand{\vs}[1]{\ensuremath{\vspace{#1}}}
    \newcommand{\wt}[1]{\ensuremath{\widetilde{#1}}}

    \newcommand{\ba}{\ensuremath{\bm{a}}}

    \newcommand{\bG}{\ensuremath{\bm{G}}}
    \newcommand{\bi}{\ensuremath{\bm{i}}}

    \newcommand{\Bm}{\ensuremath{\bm{m}}}

    \newcommand{\bS}{\ensuremath{\bm{S}}}
    \newcommand{\bt}{\ensuremath{\bm{t}}}

    \newcommand{\bx}{\ensuremath{\bm{x}}}
    \newcommand{\bX}{\ensuremath{\bm{X}}}

    \newcommand{\A}{\ensuremath{\mathcal{A}}}
    \newcommand{\C}{\ensuremath{\mathcal{C}}}
    \newcommand{\Do}{\ensuremath{\mathcal{D}}}

    \newcommand{\G}{\ensuremath{\mathcal{G}}}
    \newcommand{\I}{\ensuremath{\mathcal{I}}}
    
    \newcommand{\M}{\ensuremath{\mathcal{M}}}
    \newcommand{\N}{\ensuremath{\mathcal{N}}}

    \newcommand{\R}{\ensuremath{\mathcal{R}}}
    \newcommand{\mS}{\ensuremath{\mathcal{S}}}
    \newcommand{\Rl}{\ensuremath{\mathbb{R}}}

		\newcommand{\al}{\ensuremath{\alpha}}
    \newcommand{\Be}{\ensuremath{\beta}}

    \newcommand{\la}{\ensuremath{\lambda}}

    \newcommand{\bpi}{\ensuremath{\bm{\pi}}}

\begin{document}

    \title {Local public good provisioning in networks: \\
            A Nash implementation mechanism}
    \author{Shrutivandana~Sharma\thanks{Shrutivandana~Sharma obtained her Ph.D. from the University of Michigan, Ann Arbor (email: shrutivandana.um@gmail.com).} 
            and 
            Demosthenis~Teneketzis\thanks{Demosthenis~Teneketzis is with the University of Michigan, Ann Arbor (email: teneket@eecs.umich.edu).} %, ~\IEEEmembership{Fellow,~IEEE}}
            }
            %\thanks{ Manuscript submitted February 22, 2007.
            %Shrutivandana Sharma is a graduate student in the Department of Electrical Engineering and Computer Science,
            %University of Michigan, Ann Arbor, MI 48109-2122, USA (phone: +1-734-709-7621, fax: +1-734-763-8041, email:
            %svandana@eecs.umich.edu),
            %Demosthenis Teneketzis is with the Department of Electrical Engineering and Computer Science,
            %University of Michigan, Ann Arbor, MI 48109-2122, USA (email: teneket@eecs.umich.edu)}}
    %\author{%\authorblockN{Shrutivandana~Sharma, Demosthenis~Teneketzis,~\I{Fellow,~IEEE}}\\
%            \authorblockN{Shrutivandana~Sharma and Demosthenis~Teneketzis}\\
%            \authorblockA{
%            Department of Electrical Engineering and Computer Science, \\
%            University of Michigan, Ann Arbor, MI 48109-2122, USA \\
%            Email: svandana@umich.edu, teneket@eecs.umich.edu}}
    \maketitle

\bibliographystyle{IEEEtran}

%\mainmatter              % start of the contribution
%%
%\title{Local public good provision in networks: \\
%            A Nash implementation mechanism}
%%
%\titlerunning{Local public good provision in networks}  
%                                      % abbreviated title (for running head)
%%                                     also used for the TOC unless
%%                                     \toctitle is used
%%
%\author{Shrutivandana~Sharma\inst{1} \and Demosthenis~Teneketzis\inst{2}}
%%
%\authorrunning{Sharma et al.}   % abbreviated author list (for running head)
%%
%%%%% list of authors for the TOC (use if author list has to be modified)
%\tocauthor{Shrutivandana~Sharma, Demosthenis~Teneketzis}
%%
%\institute{Electrical Engineering and Computer Science\\
%           University of Michigan, Ann Arbor, MI 48109, USA \\
%					\email{svandana@umich.edu},
%      \and
%					 Electrical Engineering and Computer Science\\
%           University of Michigan, Ann Arbor, MI 48109, USA \\
%					\email{teneket@eecs.umich.edu} \\
%					\url{http://www.eecs.umich.edu/~teneket}}
%
%\maketitle              % typeset the title of the contribution
%% \index{Ekeland, Ivar} % entries for the author index
%% \index{Temam, Roger}  % of the whole volume
%% \index{Dean, Jeffrey}
%
%\bibliographystyle{abbrv}

\begin{abstract}        % give a summary of your paper
	In this paper we study resource allocation in decentralized information local public good networks. A network is a local public good network if each user's actions directly affect the utility of an arbitrary subset of network users. We consider networks where each user knows only that part of the network that either affects or is affected by it. Furthermore, each user's utility and action space are its private information, and each user is a self utility maximizer. This network model is motivated by several applications including wireless communications. %\cite{Sharma_thesis}. 
	For this network model we formulate a decentralized resource allocation problem and develop a decentralized resource allocation mechanism (game form)  that possesses the following properties: (i) All Nash equilibria of the game induced by the mechanism result in allocations that are optimal solutions of the corresponding centralized resource allocation problem (Nash implementation). (ii) All users voluntarily participate in the allocation process specified by the mechanism (individual rationality). (iii) The mechanism results in budget balance at all Nash equilibria and off equilibrium.
	
	% please supply keywords within your abstract
%\keywords {network, local public good, decentralized resource allocation, mechanism design, Nash implementation, budget balance, individual rationality}

\end{abstract}
%

%%%%%%%%%%%%%%%%%%%%%%%%%%%%%%%%%%%%%%%%%%%%%%%%%%%%%%%%%%%%%%%%%%%%%%%%%%%%%%%%%%%%%%%%%%%
%%%%%%%%%%%%%%%%%%%%%%%%%%%%%%%%%%%%%%%%%%%%%%%%%%%%%%%%%%%%%%%%%%%%%%%%%%%%%%%%%%%%%%%%%%%

\section{Introduction} \label{Sec_intro}

In networks individuals' actions often influence the performance of their directly connected neighbors. Such an influence of individuals' actions on their neighbors' performance can propagate through the network affecting the performance of the entire network. Examples include several real world networks. As an instance, in a wireless cellular network, the transmission of the base station to a given user (an action corresponding to this user) creates interference to the reception of other users and affects their performance. In an urban network, when a jurisdiction institutes a pollution abatement program, the benefits also accrue to nearby communities. 
%In online advertising, the utility (users' attention) that an advertisement gets may be increased or decreased due to adjacent advertisements on the webpage. 
The influence of neighbors is also observed in the spread of information and innovation in social and research networks. Networks with above characteristics are called local public good networks.

	A local public good network differs from a typical public good system in that a local public good (alternatively, the action of an individual) is accessible to and directly influences the utilities of individuals in a particular neighborhood within a big network. On the other hand a public good is accessible to and directly influences the utilities of all individuals in the system (\cite[Chapter 11]{Mas-Colell}). Because of the localized interactions of individuals, in local public good networks (such as ones described above) the information about the network is often localized; i.e., the individuals are usually aware of only their neighborhoods and not the entire network. In many situations the individuals also have some private information about the network or their own characteristics that are not known to anybody else in the network. Furthermore, the individuals may also be selfish who care only about their own benefit in the network. Such a decentralized information local public good network with selfish users gives rise to several research issues. In the next section we provide a literature survey on prior research in local public good networks.

%%%%%%%%%%%%%%%%%%%%%%%%%%%%%%%%%%%%%%%%%%%%%%%%%%%%%%%%%%%%%%%%%%

\subsection{Literature survey} \label{chap5_Subsec_lit_surv}

	There exists a large literature on local public goods within the context of local public good provisioning by various municipalities that follows the seminal work of \cite{Tiebout_56}. These works mainly consider network formation problems in which individuals choose where to locate based on their knowledge of the revenue and expenditure patterns (on local public goods) of various municipalities. In this paper we consider the problem of determining the levels of local public goods (actions of network agents) for a given network; thus, the problem addressed in this paper is distinctly different from those in the above literature. Recently, Bramoull$\acute{e}$ and Kranton \cite{Kranton_07} and Yuan \cite{Yuan_09} analyzed the influence of selfish users' behavior on the provision of local public goods in networks with fixed links among the users. The authors of \cite{Kranton_07} study a network model in which each user's payoff equals its benefit from the sum of efforts (treated as local public goods) of its neighbors less a cost for exerting its own effort. For concave benefit and linear costs, the authors analyze Nash equilibria (NE) of the game where each user's strategy is to choose its effort level that maximizes its own payoff from the provisioning of local public goods.	The authors show that at such NE, \emph{specialization} can occur in local public goods provisioning. Specialization means that only a subset of individuals contribute to the local public goods and others free ride. The authors also show that specialization can benefit the society as a whole because among all NE, the ones that are ``specialized'' result in the highest social welfare (sum of all users' payoffs). However, it is shown in \cite{Kranton_07} that none of the NE of abovementioned game can result in a local public goods provisioning that achieves the maximum possible social welfare. In \cite{Yuan_09} the work of \cite{Kranton_07} is extended to directed networks where the externality effects of users' efforts on each others' payoffs can be unidirectional or bidirectional. The authors of \cite{Yuan_09} investigate the relation between the structure of directed networks and the existence and nature of Nash equilibria of users' effort levels in those networks. For that matter they introduce a notion of ergodic stability to study the influence of perturbation of users' equilibrium efforts on the stability of NE. However, none of the NE of the game analyzed in \cite{Yuan_09} result in a local public goods provisioning that achieves optimum social welfare.

	In this paper we consider a generalization of the network models investigated in \cite{Kranton_07, Yuan_09}. Specifically, we consider a fixed network where the actions of each user directly affect the utilities of an arbitrary subset of network users. In our model, each user's utility from its neighbors' actions is an arbitrary concave function of its neighbors' action profile. Each user in our model knows only that part of the network that either affects or is affected by it. Furthermore, each user's utility and action space are its private information, and each user is a self utility maximizer. Even though the network model we consider has similarities with those investigated in \cite{Kranton_07, Yuan_09}, the problem of local public goods provisioning we formulate in this paper is different from those in \cite{Kranton_07, Yuan_09}. Specifically, we formulate a problem of local public goods provisioning in the framework of implementation theory~\fn{Refer to \cite{Jack01, Tudor05, Sharma:2011} and \cite[Chapter 3]{Sharma_thesis} and  for an introduction to implementation theory.} and address questions such as --  How should the network users communicate so as to preserve their private information, yet make it possible to determine actions that achieve optimum social welfare?  How to provide incentives to the selfish users to take actions that optimize the social welfare? How to make the selfish users voluntarily participate in any action determination mechanism that aims to optimize the social welfare? In a nutshell, the prior work of \cite{Kranton_07, Yuan_09} analyzed specific games, with linear cost functions, for local public good provision, whereas our work focusses on \emph{designing a mechanism that can induce, via nonlinear tax functions, ``appropriate'' games among the network users so as to implement the optimum social welfare in NE}.  It is this difference in the tax functions that distinguishes our results from those of \cite{Kranton_07, Yuan_09}.

	Previous works on implementation approach (Nash implementation) for (pure) public goods can be found in \cite{Groves_led_77, Hurwicz_79, Walker_81, Yan_02}. For our work, we obtained inspiration from \cite{Hurwicz_79}. In \cite{Hurwicz_79} Hurwicz presents a Nash implementation mechanism that implements the Lindahl allocation (optimum social welfare) for a public good economy. Hurwicz' mechanism also possesses the properties of individual rationality (i.e. it induces the selfish users to voluntarily participate in the mechanism) and budget balance (i.e. it balances the flow of money in the system). A local public good network can be thought of as a limiting case of a public good network, in which the influence of each public good tends to vanish on a subset of network users. However, taking the corresponding limits in the Hurwicz' mechanism does not result in a local public good provisioning mechanism with all the original properties of the Hurwicz' mechanism. In particular, such a limiting mechanism does not retain the budget balance property which is very important to avoid any scarcity/wastage of money. In this paper we address the problem of designing a local public good provisioning mechanism that possesses the desirable properties of budget balance, individual rationality, and Nash implementation of optimum social welfare. The mechanism we develop is more general than Hurwicz' mechanism; Hurwicz' mechanism can be obtained as a special case of our mechanism by setting $\R_i = \C_j = \N \fa i,j,\in\N$ (the special case where all users' actions affect all users' utilities) in our mechanism. 
	%For large scale networks where the number of users affecting a given user may be very small compared to the total number number of users in the network, 
	Our mechanism also proivdes a more efficient way to achieve the properties of Nash implementation, individual rationality, and budget balance as it uses, in general, a much smaller message space than Hurwicz' mechanism. 
	To the best of our knowledge the resource allocation problem and its solution that we present in this paper is the first attempt to analyze a local public goods network model in the framework of implementation theory. Below we state our contributions.

%%%%%%%%%%%%%%%%%%%%%%%%%%%%%%%%%%%%%%%%%%%%%%%%%%%%%%%%%%%%%%%%%%%%%%%%%%%%%%%%%%%%%%%%%%%%%%%

\subsection{Contribution of the paper} \label{chap5_Subsec_contr_chap}

    The key contributions of this paper are:
    \begin{inparaenum}
      \item  The formulation of a problem of local public goods provisioning in the framework of implementation theory.            
      \item The specification of a game form~\fn{See \cite[Chapter 3]{Sharma_thesis} and \cite{Sharma:2011, Tudor05, Jack01} for the definition of ``game form''.} (decentralized mechanism) for the above problem that,
            \begin{inparaenum}
            \item[(i)] implements in NE the optimal solution of the corresponding centralized local public good provisioning problem;
            \item[(ii)] is individually rational;~\fn{Refer to \cite[Chapter 3]{Sharma_thesis} and \cite{Sharma:2011} for the definition of ``individual rationality'' and ``implementation in NE.''} and
            \item[(iii)] results in budget balance at all NE and off equilibrium.
            \end{inparaenum}
      \end{inparaenum}

The rest of the paper is organized as follows. In Section~\ref{chap5_Subsec_mdl} we present the model of local public good network. In Section~\ref{chap5_Subsec_res_alloc_prb} we formulate the local public good provisioning problem. In Section~\ref{chap5_Subsec_game_form} we present a game form for this problem and discuss its properties in Section~\ref{chap5_Subsec_Thm}. We conclude in Section~\ref{chap5_Sec_conc} with a discussion on future directions. \\
%%
%%%%%%%%%%%%%%%%%%%%%%%%%%%%%%%%%%%%%%%%%%%%%%%%%%%%%%%%%%%%%%%%%%
%%
    {\bf Notation used in the paper:}
    We use bold font to represent vectors and normal font for scalars. We use bold uppercase letters to represent matrices. We represent the element of a vector by a subscript on the vector symbol, and the element of a matrix by double subscript on the matrix symbol. To denote the vector whose elements are all $x_i$ such that $i \in \mS$ for some set $\mS$, we use the notation $(x_i)_{i \in \mS}$ and we abbreviate it as $\bx_{\mS}$.  We treat bold {\bf 0} as a zero vector of appropriate size which is determined by the context. We use the notation $(x_i,\bm{x}^*/i)$ to represent a vector of dimension same as that of $\bm{x}^*$, whose $i$th element is $x_i$ and all other elements are the same as the corresponding elements of $\bm{x}^*$. We represent a diagonal matrix of size $N \times N$ whose diagonal entries are elements of the vector $\bx \in \Rl^N$ by $\dg{\bx}$.

%%%%%%%%%%%%%%%%%%%%%%%%%%%%%%%%%%%%%%%%%%%%%%%%%%%%%%%%%%%%%%%%%%%%%%%%%%%%%%%%%%%%%%%%%%%%%%%%%%%%%%%%%%%%%%%%%%%%%%%%%%%%%

\section{The local public good provisioning problem} \label{chap5_Sec_net_prob}

    In this section we present a model of local public good network motivated by various applications such as wireless communication \cite[Chapter 5]{Sharma_thesis}, social and information networks \cite{Yuan_09, Kranton_07}. 
   %We formulate a resource allocation problem for it. 
   We first describe the components of the model and the assumptions we make on the properties of the network. We then present a resource allocation problem for this model and formulate it as an optimization problem.
 %  ~\fn{A discussion on applciations that motivate Model~(M) can be found in \cite{Sharma_thesis}.} 

\subsection{The network model (M)} \label{chap5_Subsec_mdl}

\begin{figure}[!ht]
        \begin{minipage}[b]{0.30\linewidth}
				\centering
         \includegraphics[scale=0.30, page=1]{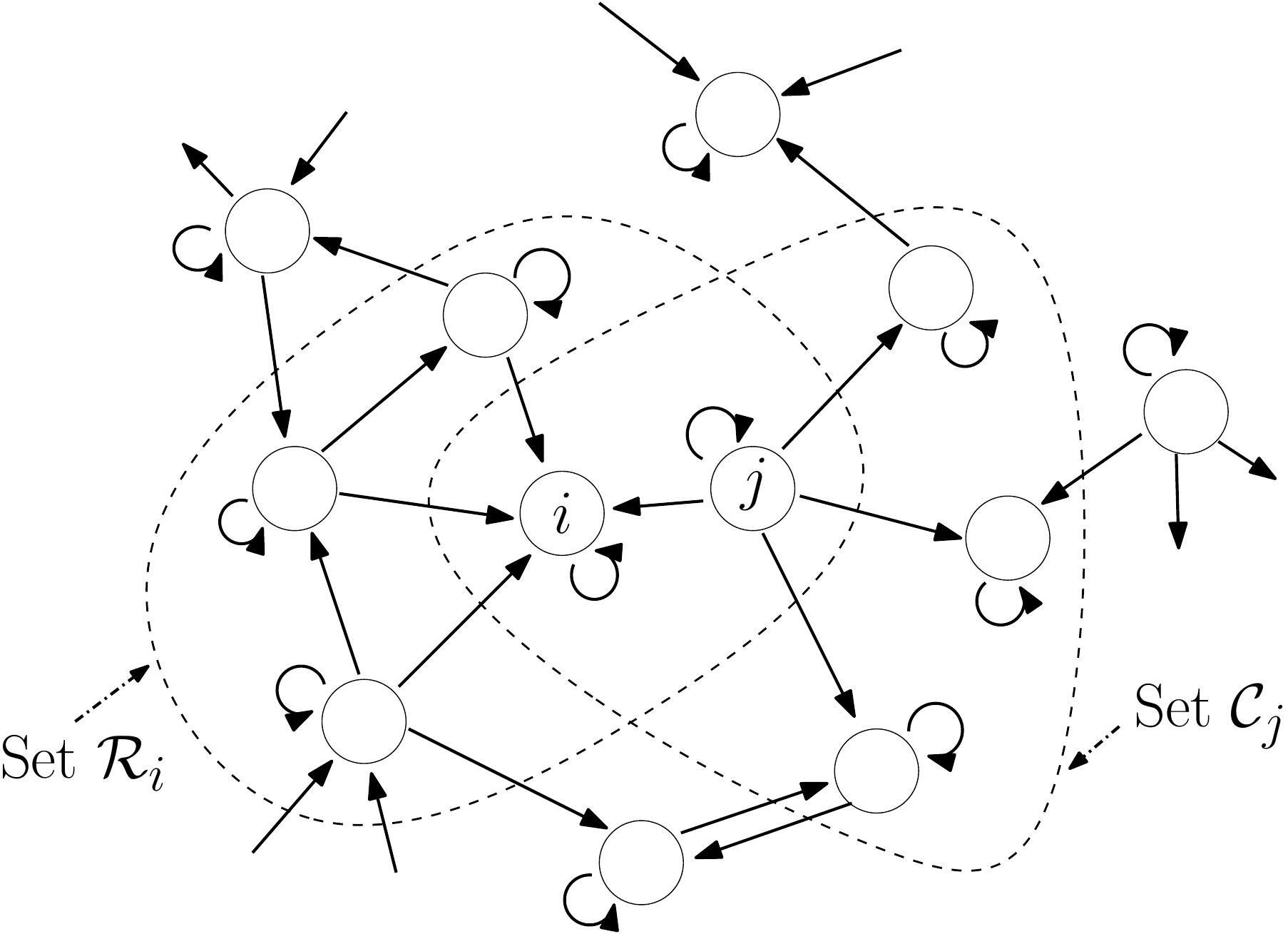}
        \caption{{\small{A local public good network depicting the 
                         Neighbor sets $\R_i$ and $\C_j$ of users $i$ and $j$ respectively.
                         }}}
        \label{chap5_Fig_large_nw}
		 	\end{minipage}
			\hspace{0.5cm}
			\begin{minipage}[b]{0.60\linewidth}
        \centering
        \includegraphics[scale=0.30, page=2]{large_network.pdf}
        \caption{Illustration of indexing rule for set $\C_j$ shown in Fig.~\ref{chap5_Fig_large_nw}. Index $\I_{rj}$ of user $r \in \C_j$ is indicated on the arrow directed from $j$ to $r$. The notation to denote these indices and to denote the user with a particular index is shown outside the dashed boundary demarcating the set $\C_j$.
        }
        \label{chap5_Fig_Cj_cyclic_index}
        \end{minipage}
  \end{figure}

    We consider a network consisting of $N$ users and one network operator. We denote the set of users by $\N := \{1,2,\dots,N\}$. 
    Each user $i\in\N$ has to take an action $a_i \in \A_i$ where $\A_i$ is the set that specifies user $i$'s feasible actions.  
    In a real network, a user's actions can be consumption/generation of resources or decisions regarding various tasks.     
    We assume that,
    \begin{Ass} \label{chap5_Ass_Ai_pvt}
        For all $i \in \N$, $\A_i$ is a convex and compact set in $\Rl$ that includes 0.~\fn{In this paper we assume the sets $\A_i, i \in \N$, to be in $\Rl$ for simplicity. However, the decentralized mechanism and the results we present in this paper can be easily generalized to the scenario where for each $i \in \N$, $\A_i \subset \Rl^{n_i}$ is a convex and compact set in higher dimensional space $\Rl^{n_i}$. Furthermore, each space $\Rl^{n_i}$ can be of a different dimension $n_i$ for different $i \in \N$.} Furthermore, for each user $i \in \N$, the set $\A_i$ is its private information, i.e. $\A_i$ is known only to user $i$ and nobody else in the network.
    \end{Ass}
    Because of the users' interactions in the network, the actions taken by a user directly affect the performance of other users in the network. Thus, the performance of the network is determined by the collective actions of all users. We assume that the network is large-scale, therefore,  every user's actions directly affect only a subset of network users in $\N$. Thus we can treat each user's action as a local public good. We depict the above feature by a directed graph as shown in Fig.~\ref{chap5_Fig_large_nw}. In the graph, an arrow from $j$ to $i$ indicates that user $j$ affects user $i$; we represent the same in the text as $j \raro i$. We assume that $i \raro i$ for all $i \in \N$.
%    \begin{figure}[!ht]
%        \centering
%        \includegraphics[scale=0.30, page=1]{large_network.pdf}
%        \caption{{\small{A local public good network depicting the 
%                         Neighbor sets $\R_i$ and $\C_j$ of users $i$ and $j$ respectively.
%                         }}}
%        \label{chap5_Fig_large_nw}
%    \end{figure}
%%
%    \begin{figure}[!ht]
%        \centering
%        \includegraphics[scale=0.30, page=2]{large_network.pdf}
%        \caption{Illustration of indexing rule for set $\C_j$ shown in Fig.~\ref{chap5_Fig_large_nw}. Index $\I_{rj}$ of user $r \in \C_j$ is indicated on the arrow directed from $j$ to $r$. The notation to denote these indices and to denote the user with a particular index is shown outside the dashed boundary demarcating the set $\C_j$.
%        }
%        \label{chap5_Fig_Cj_cyclic_index}
%    \end{figure}
%%{\bf Note: In this draft Fig. 1 appears on page 31.}
    Mathematically, we denote the set of users that affect user $i$ by $\R_i := \{k \in \N \mid k \raro i\}$. Similarly, we denote the set of users that are affected by user $j$ by $\C_j := \{k \in \N \mid j \raro k\}$. We represent the interactions of all network users together by a graph matrix $\bG := [G_{ij}]_{N\times N}$. The matrix $\bG$ consists of 0's and 1's,  where $G_{ij} = 1$ represents that user $i$ is affected by user $j$, i.e. $j \in \R_i$ and $G_{ij} = 0$ represents no influence of user $j$ on user $i$, i.e. $j \nin \R_i$. Note that $\bG$ need not be a symmetric matrix. Because $i \raro i$, $G_{ii} = 1$ for all $i \in \N$. We assume that,
    \begin{Ass} \label{chap5_Ass_G_indep_a}
        The sets $\R_i, \C_i, i \in \N$, are independent of the users' action profile $\ba_{\N} := (a_k)_{k \in \N} \in \prod_{k \in \N} \A_k$. Furthermore, for each $i \in \N$, $|\C_i| \geq 3$. 
    \end{Ass}
        We consider the condition $|\mathcal{C}_i| \geq 3, i\in\N,$ so as to ensure construction of a mechanism that is budget balanced at all possible allocations, those that correspond to Nash equilibria as well as those that correspond to off-equilibrium messages.
    For examples of applications where Assumption~\ref{chap5_Ass_G_indep_a} holds, see 
    %\cite[Chapter 5]{Sharma_thesis} 
    Section~\ref{chap5_Subsec_apps}
    and \cite{Yuan_09, Kranton_07}.

    We assume that,
    \begin{Ass}\label{Ass_user_i_knows_superset_aj}
    	Each user $i\in\N$ knows that the set of feasible actions $\A_j$ of each of its neighbors $j\in\R_i$ is a convex and compact subset of $\Rl$ that includes 0.   
    \end{Ass}
    
    The performance of a user that results from actions taken by the users affecting it is quantified by a utility function. We denote the utility of user $i \in \N$ resulting from the action profile $\ba_{\R_i} := (a_k)_{k \in \R_i}$ by $u_i(\ba_{\R_i})$. We assume that,
    \begin{Ass} \label{chap5_Ass_util_conc_pvt}
        For all $i \in \N$, the utility function $u_i : \Rl^{|\R_i|} \raro \Rl \cup \{-\infty\}$ is concave in $\ba_{\R_i}$ and $u_i(\ba_{\R_i}) = -\infty$ for $a_i \nin \A_i$.~\fn{Note that $a_i$ is always an element of $\ba_{\R_i}$ because $i \raro i$ and hence $i \in \R_i$.} The function $u_i$ is user $i$'s private information.
    \end{Ass}
    The assumptions that $u_i$ is concave and is user $i$'s private information are motivated by applications described in 
    %\cite[Chapter 5]{Sharma_thesis} 
    Section~\ref{chap5_Subsec_apps}
    and \cite{Yuan_09, Kranton_07}. The assumption that $u_i(\ba_{\R_i}) = -\infty$ for $a_i \nin \A_i$ captures the fact that an action profile $(\ba_{\R_i})$ is of no significance to user  $i$ if $a_i \nin \A_i$. 
    We assume that,
    \begin{Ass} \label{chap5_Ass_selfish_users}
        Each network user $i \in \N$ is selfish, non-cooperative, and strategic.
    \end{Ass}
    Assumption~\ref{chap5_Ass_selfish_users} implies that the users have an incentive to misrepresent their private information, e.g. a user $i\in\N$ may not want to report to other users or to the network operator its true preference for the users' actions, if this results in an action profile in its own favor.

    Each user $i \in \N$ pays a tax $t_i \in \Rl$ to the network operator. This tax can be imposed for the following reasons: (i) For the use of the network by the users. (ii) To provide incentives to the users to take actions that achieve a network-wide performance objective. The tax is set  according to the rules specified by a mechanism and it can be either positive or negative for a user. With the flexibility of either charging a user (positive tax) or paying compensation/subsidy (negative tax) to a user, it is possible to induce the users to behave in a way such that a network-wide performance objective is achieved. For example, in a network with limited resources, we can set ``positive tax'' for the users that receive resources close to the amounts requested by them and we can pay  ``compensation'' to the users that receive resources that are not close to their desirable ones. Thus, with the available resources, we can satisfy all the users and induce them to behave in a way that leads to a resource allocation that is optimal according to a network-wide performance criterion. 
    We assume that,
    \begin{Ass} \label{chap5_Ass_net_op_acct}
        The network operator does not have any utility associated with the users' actions or taxes. It does not derive any profit from the users' taxes and acts like an accountant that redistributes the tax among the users according to the specifications of the allocation mechanism.
    \end{Ass}
    Assumption~\ref{chap5_Ass_net_op_acct} implies that the tax is charged in a way such that
    \begin{equation}
        \sum_{i \in \N}  t_i  = 0.   \label{chap5_Eq_sum_ti_0}
    \end{equation}
    To describe the ``overall satisfaction'' of a user from the performance it receives from all users' actions and the tax it pays for it, we define an ``aggregate utility function'' $u_i^A(\ba_{\R_i}, t_i): \Rl^{|\R_i| + 1} \raro \Rl \cup \{-\infty\}$ for each user $i \in \N$:
    \begin{equation}
    \begin{split}
         u_i^A(\ba_{\R_i}, t_i) := 
             \left\{ \begin{array}{rl} 
                    -t_i + u_i(\ba_{\R_i}), &  \mb{if} \ a_i \in  \A_i, 
                           a_j \in \Rl, j \in \R_i \bs \{i\}, \\
                    -\infty, &  \mb{otherwise.}
                    \end{array} \right. 
    \end{split} \label{chap5_Eq_u_i^A}
    \end{equation}
    Because $u_i$ and $\A_i$ are user $i$'s private information (Assumptions~\ref{chap5_Ass_Ai_pvt} and \ref{chap5_Ass_util_conc_pvt}), the aggregate utility $u_i^A$ is also user $i$'s private information. As stated in Assumption~\ref{chap5_Ass_selfish_users}, users are non-cooperative and selfish. Therefore, \emph{the users are self aggregate utility maximizers}.

    In this paper we restrict attention to static problems, i.e. we assume,
    \begin{Ass} \label{chap5_Ass_net_const}
        The set $\N$ of users, the graph $\bG$, users' action spaces $\A_i, i \in \N$, and their utility functions $u_i, i \in \N$, are fixed in advance and they do not change during the time period of interest.
    \end{Ass}

    We also assume that,
    \begin{Ass} \label{chap5_Ass_nbr_know}
        Every user $i \in \N$ knows the set $\R_i$ of users that affect it as well as the set $\C_i$ of users that are affected by it. The network operator knows $\R_i$ and $\C_i$ for all $i \in \N$.
    \end{Ass}
    In networks where the sets $\R_i$ and $\C_i$ are not known to the users beforehand, Assumption~\ref{chap5_Ass_nbr_know} is still reasonable because of the following reason. As the graph $\bG$ does not change during the time period of interest (Assumption~\ref{chap5_Ass_net_const}), the information about the neighbor sets $\R_i$ and $\C_i, i \in \N$, can be passed to the respective users by the network operator before the users determine their actions. Alternatively, the users can themselves determine the set of their neighbors before determining their actions.~\fn{The exact method by which the users get information about their neighbor sets in a real network depends on the network characteristics.}
    Thus, Assumption~\ref{chap5_Ass_nbr_know} can hold true for the rest of the action determination process.
%%%%
%		In the next section we present a local public good provisioning problem for Model~(M).
 In the next section we present some applications that motivate Model~(M).
 
%%%%%%%%%%%%%%%%%%%%%%%%%%%%%%%%%%%%%%%%%%%%%%%%%%%%%%%%%%%%
\subsection{Applications} \label{chap5_Subsec_apps}
%%%%%%%%%%%%%%%%%%%%%%%%%%%%%%%%%%%%%%%%%%%%%%%%%%%%%%%%%%%%
%%%%%%%%%%%%%%%%%%%%%%%%%%%%%%%%%%%%%%%%%%%%%%%%%%%%%%%%%%%%
\subsubsection{Application A: Power allocation in cellular networks} \label{chap5_Subsubsec_power_alloc}
%%%%%%%%%%%%%%%%%%%%%%%%%%%%%%%%%%%%%%%%%%%%%%%%%%%%%%%%%%%%

\begin{figure}[!ht]
        \begin{minipage}[b]{0.30\linewidth}
				\centering
         \includegraphics[scale=.25,page=1]{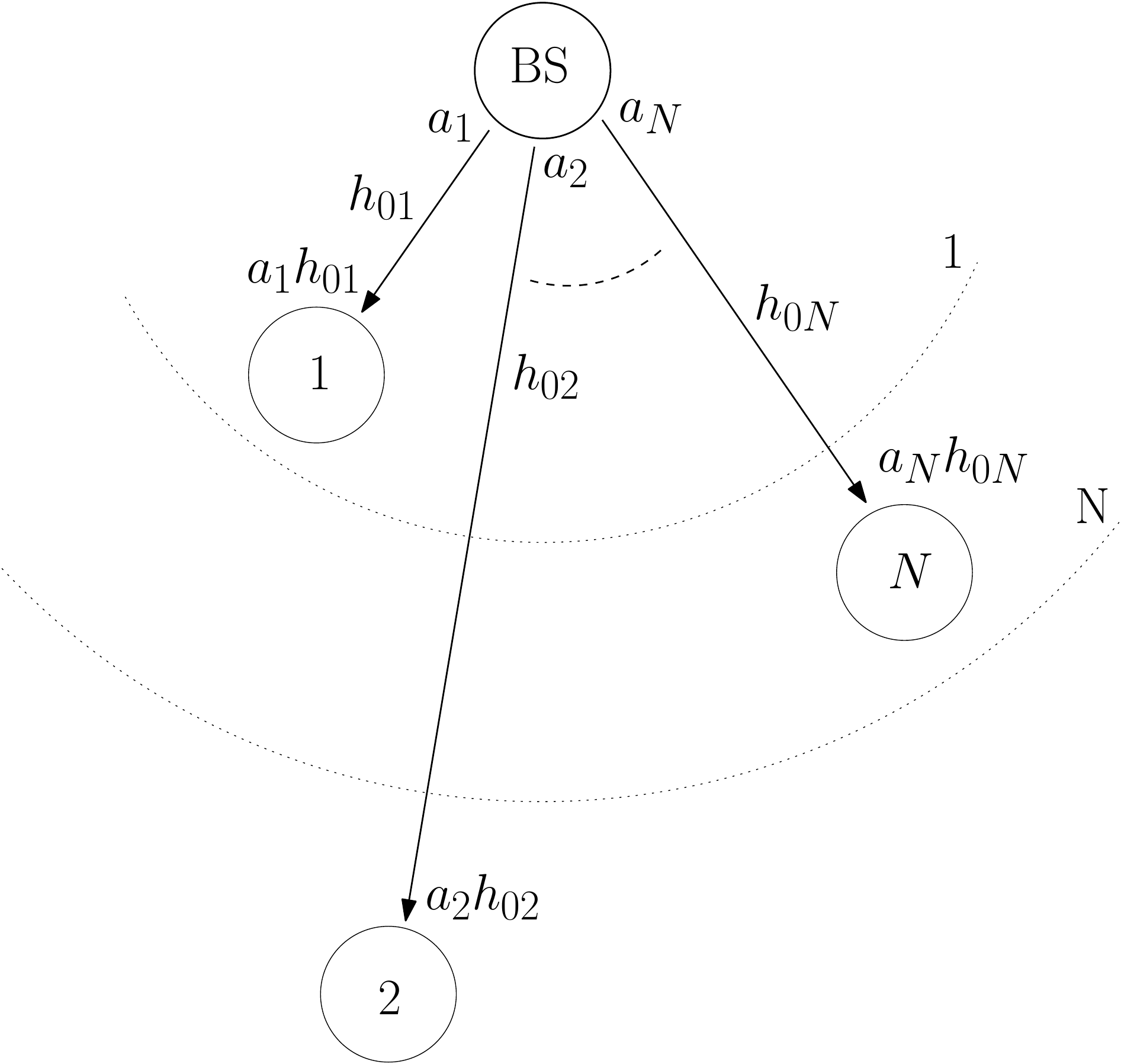}
        \caption{A downlink network with $N$ mobile users and one base station}
        \label{chap5_Fig_down_link}
		 	\end{minipage}
			\hspace{0.5cm}
			\begin{minipage}[b]{0.55\linewidth}
        \centering
        \includegraphics[scale=.30,page=1]{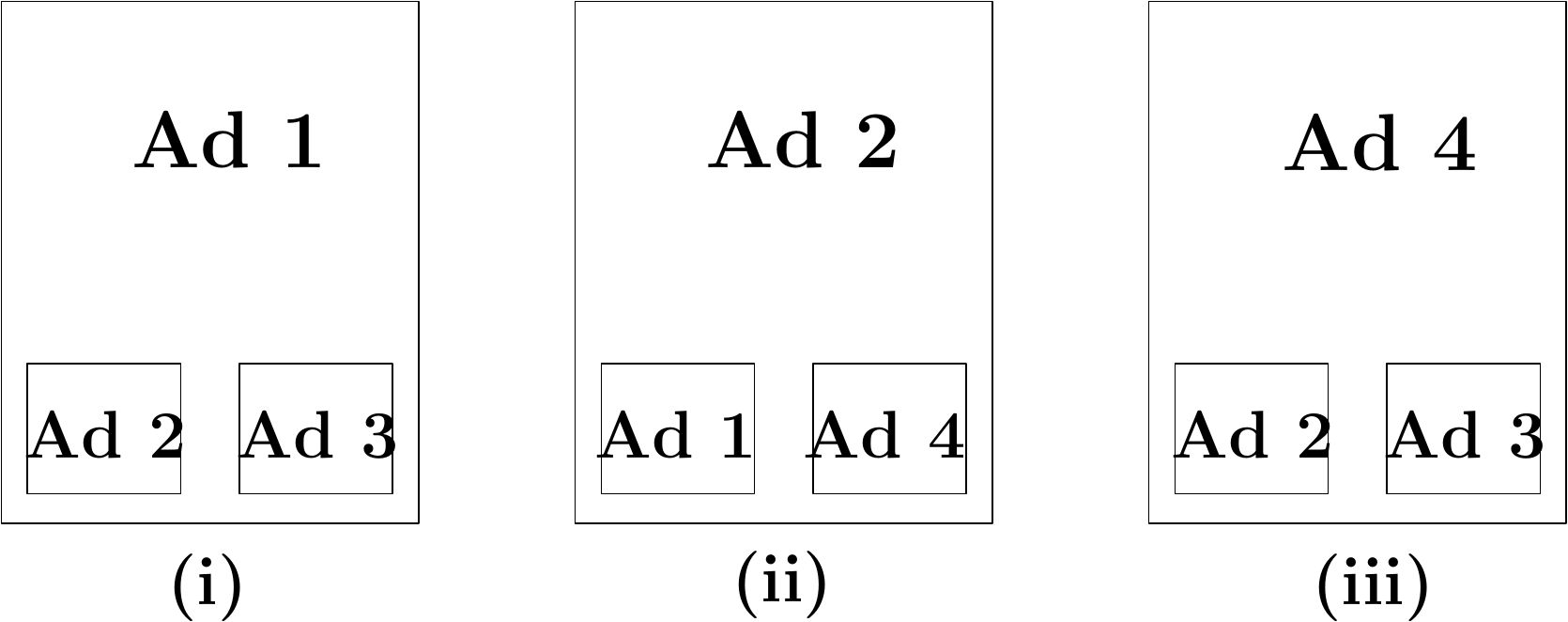}
         \caption{{\small{Three display ad clusters, each consisting of one main ad and two sub ads.}}}
        \label{chap5_Fig_ad_cluster}
        \end{minipage}
  \end{figure}

    Consider a single cell downlink wireless data network consisting of a Base Station (BS) and $N$ mobile users as shown in Fig.~\ref{chap5_Fig_down_link}.
    %%%
	%%% INCLUDE
%    \begin{figure}[!ht]
%        \centering
%        \includegraphics[scale=.40,page=1]{Figs/ad_cluster.pdf}
%        \caption{{\small{Three display ad clusters, each consisting of one main ad and two sub ads.}}}
%        \label{chap5_Fig_ad_cluster}
%    \end{figure}
%%%
%%%
%    %%%% INCLUDE
%    \begin{figure}[!ht]
%       \centering
%       \includegraphics[scale=.3,page=1]{Figs/cell_net.pdf}
%       \caption{A downlink network with $N$ mobile users and one base station}
%       \label{chap5_Fig_down_link}
%    \end{figure}
%%%
    The BS uses Code Division Multiple Access  technology (CDMA) to transmit data to the users and each mobile user uses Minimum Mean Square Error Multi-User Detector (MMSE-MUD) receiver to decode its data. The signature codes used by the BS are not completely orthogonal as this helps increase the capacity of the network. Because of non-orthogonal codes,  each user experiences interference due to the BS transmissions intended for other users. However, as the users in the cell are at different distances from the BS, and the power transmitted by the BS undergoes propagation loss, not all transmissions by the BS create interference to every user. For example, let us look at arcs $1$ and $N$ shown in Fig.~\ref{chap5_Fig_down_link} that are centered at the BS. Suppose the radius of arc $1$ is much smaller than that of arc $N$. Then, the signal transmitted by the BS for users inside circle~$1$ (that corresponds to arc~$1$) will become negligible when it reaches  outside users such as user $N$ or user $2$. On the other hand, the BS signals transmitted for user $N$ and user $2$ will be received with significant power by the users inside circle~$1$. This asymmetric interference relation between the mobile users can be depicted in a graph similar to one shown in Fig.~\ref{chap5_Fig_large_nw}. In the graph an arrow from $j$ to $i$ would represent that the signal transmitted for user $j$  also affects user $i$. Note that since the signal transmitted for user $i$ must reach $i$, the assumption $i \raro i$ made in Section~\ref{chap5_Subsec_mdl} holds in this case. If the users do not move very fast in the network, the network topology can be assumed to be fixed for small time periods. Therefore, if the BS transmits some pilot signals to all network users, the users can figure out which signals are creating interference to their signal reception. Thus, each user would know its (interfering) neighbor set as assumed in Assumption~\ref{chap5_Ass_nbr_know}. Note that if the power transmitted by the BS to the users change, it may result in a change in the set of interfering neighbors of each user. This is different from Assumption~\ref{chap5_Ass_G_indep_a} in Model~(M). However, if the transmission power fluctuations resulting from a power allocation mechanism are not large, the set of interfering neighbors can be assumed to be fixed, and this can be approximated by Assumption~\ref{chap5_Ass_G_indep_a}.

    The Quality of Service (QoS) that a user receives from decoding its data is quantified by a utility function. Due to interference the utility $u_i(\cdot)$ of user $i, i \in \N$, is a function of the vector $\ba_{\R_i}$, where $a_j$ is the transmission power used by the BS to transmit signals to user $j, j \in \N$, and $\R_i$ is the set of users such that the signals transmitted by the BS to users in $\R_i$ also reach user $i$. Note that in this case all transmissions, in other words the actions $a_i, i \in \N,$ are carried out by the BS unlike Model~(M) where each user $i \in \N$ takes its own action $a_i$. However, as we discuss below, the BS is only an agent which executes the outcome of the mechanism that determines these transmission powers.   Thus, we can embed the downlink network scenario into Model~(M) by treating each $a_i$ as a decision ``corresponding'' to user $i, i \in \N$, which is executed by the BS for $i$. Since each user uses an MMSE-MUD receiver, a measure of user $i$'s $(i \in \N)$ utility can be the negative of the MMSE at the output of its receiver,~\fn{See \cite{Verdu} for the derivation of (\ref{chap5_Eq_u_i_cell}).} i.e.,
\begin{equation}
 \begin{split}
         u_i(\ba_{\R_i})
    = - M\!M\!S\!E_i  
    = - \min_{\bm{z_i}^T \in \Rl^{1 \times N}}
         E[\| b_i - \bm{z_i}^T \bm{y_i} \|^2] 
     = - \big[(\bm{I} + \frac{2}{N_{0i}} \bS_i \bX_{\R_i} \bS_i)^{-1}\big]_{ii},
        \quad i \in \N.
   \label{chap5_Eq_u_i_cell}
 \end{split}
\end{equation}
    In (\ref{chap5_Eq_u_i_cell}) $b_i$ is the transmitted data symbol for user $i$, $\bm{y_i}$ is the output of user $i$'s matched filter generated from its received data, $\bm{I}$ is the identity matrix of size $N \times N$, $N_{0i}/2$ is the two sided power spectral density (PSD) of thermal noise, $\bX_{\R_i}$ is the cross-correlation matrix of signature waveforms corresponding to the users $j \in \R_i$, and $\bS_i:=\dg{(S_{ij})_{j \in \R_i}}$ is the diagonal matrix consisting of the signal amplitudes $S_{ij}, j \in \R_i$, received by user $i$.  $S_{ij}$ is related to $a_j$ as $S_{ij}^2 = a_j h_{0i}$, $j \in \R_i$, where $h_{0i}$ is the channel gain from the BS to user $i$ which represents the power loss along this path. As shown in \cite{Sharma:2009, Sharma:2011}, the utility function given by (\ref{chap5_Eq_u_i_cell}) is close to concave in $\ba_{\R_i}$. Thus, Assumption~\ref{chap5_Ass_util_conc_pvt} in Model~(M) can be thought of as an approximation to the downlink network scenario.

    Note that to compute user $i$'s utility given in (\ref{chap5_Eq_u_i_cell}), knowledge of $N_{0i}$, $\bX_{\R_i}$, and $h_{0i}$ is required. The BS knows $\bX_{\R_i}$ for each $i \in \N$ as it selects the signature waveform for each user. On the other hand, user $i, i\in\N$, knows the PSD $N_{0i}$ of thermal noise and the channel gain $h_{0i}$ as these can be measured only at the respective receiver. Consider a network where the mobile users are selfish and non cooperative. Then, these users may not want to reveal their measured values $N_{0i}$ and $h_{0i}$. On the other hand if the network operator that owns the BS does not have a utility and is not selfish, then, the BS can announce the signature waveforms it uses for each user. Thus, each user $i \in \N$ would know its corresponding cross correlation matrix $\bX_{\R_i}$ and consequently, its utility function $u_i$. However, since $N_{0i}$ and $h_{0i}$ are user $i$'s private information, the utility function $u_i$ is private information of $i$ which is similar to Assumption~\ref{chap5_Ass_util_conc_pvt} in Model~(M). If the wireless channel conditions vary slowly compared to the time period of interest, the channel gains and hence the users' utility functions can be assumed to be fixed. As mentioned earlier, for slowly moving users the network topology and hence the set of interfering neighbors can also be assumed to be fixed. These features are captured by Assumption~\ref{chap5_Ass_net_const} in Model~(M).

    In the presence of limited resources, the provision of desired QoS to all network users may not be possible. To manage the provision of QoS under such a situation the network operator (BS) can charge tax to the users and offer them the following tradeoff. It charges positive tax to the users that obtain a QoS close to their desirable one, and compensates the loss in the QoS of other users by providing a subsidy to them. Such a redistribution of money among users through the BS is possible under Assumption~\ref{chap5_Ass_net_op_acct} in Model~(M).
%
%
%%%%%%%%%%%%%%%%%%%%%%%%%%%%%%%%%%%%%%%%%%%%%%%%%%%%%%%%%%%%
\subsubsection{Application B: Online advertising} \label{Subsubsec_onl_adv}
%%%%%%%%%%%%%%%%%%%%%%%%%%%%%%%%%%%%%%%%%%%%%%%%%%%%%%%%%%%%

	Consider an online guaranteed display (GD) ad system. In existing GD systems, individual advertisers sign contracts with web-publishers in which web-publishers agree to serve (within some given time period) a fixed number of impressions~\fn{A display instance of an ad} of each ad for a lump sum payment. Here we consider an extension of current GD systems in which multiple advertisers can form clusters as shown in Fig.~\ref{chap5_Fig_ad_cluster} and contracts can be signed for the number of impressions of each cluster.  
	Figure~\ref{chap5_Fig_ad_cluster} shows three display ad clusters. In each ad cluster there is one main ad and two sub ads. For example in Fig.~\ref{chap5_Fig_ad_cluster}-(i), ad~1 is the main ad and ad~2 and ad~3 are the two sub ads. 
	Suppose that the ad clusters are formed in a way so that in each cluster, the main ad creates a positive externality~\fn{See \cite[Chapter 11]{Mas-Colell} for the definition of externality.} to the sub ads. For example, ad~1 can be a Honda ad, whereas ad~2 and ad~3 can be ads of local Honda dealer and local Honda mechanic. We call the ad cluster in which ad $i$ appears as the main ad as cluster $i$.~\fn{We assume that there is at most one ad cluster in which ad $i$ appears as the main ad; hence such a notation is well defined.} The arrangements of ads in clusters can be described by a graph similar to one shown in Fig.~\ref{chap5_Fig_large_nw}. In this graph an arrow from $j$ to $i$ would represent that ad $i$ appears in cluster $j$. For each impression of cluster $i$, $i \in \{1,2,4\}$, advertiser $i$ pays some fixed prespecified amount of money (bid) $b_i \in \Rl_+$ to the web publisher. Suppose the bid $b_i$ is known only to advertiser $i$ and the web publisher. Let $a_i, i \in \{1,2,4\},$ denote the number of impressions of clusters $i$ delivered to the users. Furthermore, let $A_i^{max}$ be the maximum number of impressions advertiser $i$ can request for cluster $i$, i.e.  $0 \leq a_i \leq A_i^{max}$. The constraint $A_i^{max}$ may arise due to the budget constraint of advertiser $i$, or due to the restrictions imposed by the web publisher. For these reasons $A_i^{max}$ may be private information of advertiser $i$ (similar to Assumption~\ref{chap5_Ass_Ai_pvt}) or private knowledge between advertiser $i$ and the web publisher. Note that in the ad network, the number of impressions $a_i, i \in \{1,2,4\},$ can take only natural number values; therefore, the assumption of convex action sets $\A_i, i \in \N,$ in Assumption~\ref{chap5_Ass_Ai_pvt} can be thought of as an approximation to this case. Note also that in the above ad network example, no cluster is associated with advertiser 3 or the web publisher (i.e. there is no cluster with main ad 3 or an ad of the web publisher). Such a scenario can be captured by Model~(M) by associating dummy action variables $a_3$ and $a_0$ with advertiser 3 and web publisher respectively, and assuming $A_3^{max} = A_0^{max} = 0$.

	Because of the way clusters are formed, each advertiser obtains a non-negative utility from the impressions of the clusters that it is part of. Thus we can represent the utilities of the four advertisers in the ad network of Fig.~\ref{chap5_Fig_ad_cluster} as follows: 
	\begin{equation}
		\begin{split}
			u_1(a_1, a_2) & = c_{11} a_1 + c_{12} a_2 - b_1 a_1 \\
			u_2(a_1, a_2, a_4) & = c_{21} a_1 + c_{22} a_2 + c_{24} a_4 - b_2 a_2  \\
			u_3(a_1, a_4) &= c_{31} a_1 + c_{34} a_4 \\
			u_4(a_2, a_4) &= c_{42} a_2 + c_{44} a_4 - b_4 a_4.
			\label{chap5_Eq_advt_util}
		\end{split}
	\end{equation} 
	In (\ref{chap5_Eq_advt_util}) $c_{ij} \in \Rl_+, i,j \in \{1,2,3,4\}$ are non negative real valued constants. The constant $c_{ij}$ represents the value obtained by advertiser $i$ from each impression of cluster $j$. Suppose that for each $j$, $c_{ij}$ is advertiser $i$'s private information. The term $-b_i a_i, i \in \{1,2,4\}$ represents the loss in utility/monetary value incurred by advertiser $i$ due to the prespecified payment it makes to the web publisher. Because the web publisher receives payments from the advertisers, it also obtains a utility as follows:
    \begin{equation}
		\begin{split}
			u_0(a_1, a_2, a_4) & = b_1 a_1 + b_2 a_2 + b_4 a_4. 
			\label{chap5_Eq_pub_util}
		\end{split}
	\end{equation} 
	Since each bid $b_i, i \in \{1,2,4\}$, is known to the web publisher, and none of the advertisers know all of these bids, the utility function $u_0$ is web publisher's private information. Similarly, since $c_{ij}$ for each $j$ and the bid $b_i$ are private information of advertiser $i$, for each $i \in \{1,2,4\}$ the utility function $u_i$ is advertiser $i$'s private information. These properties of utility functions along with their linearity given by (\ref{chap5_Eq_advt_util}) and (\ref{chap5_Eq_pub_util}) are modeled  by Assumption~\ref{chap5_Ass_util_conc_pvt} in Model~(M). If we assume that the arrangements of ads in clusters and the bids $b_i, i \in \{1,2,4\},$ of advertisers are predetermined, and do not change with any decision regarding the impressions delivery of various clusters, then this leads to assumptions~\ref{chap5_Ass_G_indep_a} and \ref{chap5_Ass_net_const} in Model~(M).

	As represented by (\ref{chap5_Eq_advt_util}), each advertiser benefits from a number of other advertisers by being part of their ad clusters. For this reason, in addition to making a direct payment to the web publisher, each advertiser should also make a payment to those advertisers that create positive externalities to it. Furthermore, because of their mutual payments the web publisher may offer discounts to the advertisers in their direct payments to her. These discounts can indirectly encourage advertisers to participate in the clustered GD ad scheme. However, one problem in the implementation of above type of payments/discounts is that in a big network, the advertisers may not have direct contracts with all their cluster sharing advertisers. Furthermore, if the advertisers and the web publisher are strategic and self utility maximizers, they would try to negotiate payments so as to get cluster impressions that maximize their respective utilities. In such a scenario, the abovementioned distribution of money among advertisers and the web publisher can be facilitated through a third party ad agency to which all advertisers and the web publisher can subscribe. The role of an ad agency can be mapped to that of the network operator in Model~(M).
	% described by Assumption~\ref{chap5_Ass_net_op_acct}. 	

	Having discussed various applications that motivate Model (M), we now go back to this generic model and formulate a resource allocation problem for it.

%%%%%%%%%%%%%%%%%%%%%%%%%%%%%%%%%%%%%%%%%%%%%%%%%%%%%%%%%%%%%%%%%%%%%%%%%%%%%%%%%%%%%%%%%%%%%%%%%%%%%%%%%%%%%%%%%%
\subsection{The decentralized local public good provisioning problem ($P_D$)} \label{chap5_Subsec_res_alloc_prb}

    For the network model (M) we wish to develop a mechanism to determine the users' action and tax profiles $(\ba_{\N}, \bt_{\N}) := ((a_1, a_2, \dots,$ $a_N),  (t_1, t_2, \dots, t_N))$. We want the mechanism to work under the decentralized information constraints of the model and to lead to a solution to the following centralized problem. \\
    {\bf The centralized problem ($\bm{P_C}$)}
    \begin{equation}
    \begin{split}
        \max_{(\ba_{\N}, \bt_{\N})} & \quad \sum_{i \in \N} u_i^A(\ba_{\R_i}, t_i)  \\
    \mb{s.t.} & \quad \sum_{i \in \N} t_i = 0
    \label{chap5_Eq_Pc_obj1}
    \end{split}
    \end{equation}
    \begin{equation}
    \begin{split}
      \equiv \ \   &  \max_{(\ba_{\N}, \bt_{\N}) \in \Do}  \quad \sum_{i \in \N}  u_i(\ba_{\R_i})  \\
        & \mb{where},\ \  \Do :=  \{ (\ba_{\N}, \bt_{\N}) \in \Rl^{2N} \mid  a_i \in \A_i \fa i \in \N;  \sum_{i \in \N}  t_i = 0 \}
    \label{chap5_Eq_Pc_obj2}
    \end{split}
    \end{equation}

     The centralized optimization problem (\ref{chap5_Eq_Pc_obj1}) is equivalent to (\ref{chap5_Eq_Pc_obj2}) because for $(\ba_{\N}, \bt_{\N}) \nin \Do$, the objective function in (\ref{chap5_Eq_Pc_obj1}) is negative infinity by (\ref{chap5_Eq_u_i^A}). Thus $\Do$ is the set of feasible solutions of Problem~($P_C$). Since by Assumption~\ref{chap5_Ass_util_conc_pvt}, the objective function in (\ref{chap5_Eq_Pc_obj2}) is concave in $\ba_{\N}$ and the sets $\A_i, i\in\N,$ are convex and compact, there exists an optimal action profile $\ba_{\N}^*$ for Problem~($P_C$). Furthermore, since the objective function in (\ref{chap5_Eq_Pc_obj2}) does not explicitly depend on $\bt_{\N}$, an optimal solution of Problem $(P_C)$ must be of the form $(\ba_{\N}^*, \bt_{\N})$, where $\bt_{\N}$ is any feasible tax profile for Problem $(P_C)$, i.e. a tax profile that satisfies (\ref{chap5_Eq_sum_ti_0}).

     The solutions of Problem~($P_C$) are ideal action and tax profiles that we would like to obtain. If there exists an entity that has centralized information about the network, i.e. it knows all the utility functions $u_i, i\in\N$, and all action spaces $\A_i, i\in\N$, then that entity can compute the above ideal profiles by solving Problem~($P_C$). Therefore, we call the solutions of Problem~($P_C$) optimal centralized allocations. In the network described by Model~(M), there is no entity  that knows perfectly all the parameters that describe Problem~$(P_C)$ (Assumptions~\ref{chap5_Ass_Ai_pvt} and \ref{chap5_Ass_util_conc_pvt}). Therefore, we need to develop a mechanism that allows the network users to communicate with one another and that leads to optimal solutions of Problem~$(P_C)$. Since a key assumption in Model~(M) is that the users are strategic and non-cooperative, the mechanism we develop must take into account the users' strategic behavior in their communication with one another. To address all of these issues we take the approach of implementation theory \cite{Jack01} for the solution of the decentralized local public good provisioning problem for Model~(M). Henceforth we call this decentralized allocation problem as Problem ($P_D$). In the next section we present a decentralized mechanism (game form) for local public good provisioning that works under the constraints imposed by Model~(M) and achieves optimal centralized allocations.

%%%%%%%%%%%%%%%%%%%%%%%%%%%%%%%%%%%%%%%%%%%%%%%%%%%%%%%%%%%%%%%%%%%%%%%%%%%%%%%%%%%%%%%%%%%%%%%

\section{A decentralized local public good provisioning mechanism} \label{chap5_Sec_dec_mech}

    For Problem $(P_D)$, we want to develop a game form (message space and outcome function) that is \emph{individually rational}, \emph{budget balanced}, and that \emph{implements in Nash equilibria} the goal correspondence defined by the solution of Problem~$(P_C)$.~\fn{The definition of game form, goal correspondence, individual rationality, budget balance and implementation in Nash equilibria is given in \cite[Chapter 3]{Sharma_thesis}.} Individual rationality guarantees voluntary participation of the users in the allocation process specified by the game form, budget balance guarantees that there is no money left unclaimed/unallocated at the end of the allocation process (i.e. it ensures (\ref{chap5_Eq_sum_ti_0})), and implementation in NE guarantees that the allocations corresponding to the set of NE of the game induced by the game form are a subset of the optimal centralized allocations (solutions of Problem~($P_C$)).

    We would like to clarify at this point the definition of individual rationality (voluntary participation) in the context of our problem. Note that in the network model~(M), the participation/non-participation of each user determines the network structure and the set of local public goods (users' actions) affecting the participating users. To define individual rationality in this setting we consider our mechanism to be consisting of two stages as discussed in \cite[Chapter 7]{Fudenberg_91}. In the first stage, knowing the game form, each user makes a decision whether to participate in the game form or not. The users who decide not to participate are considered out of the system. Those who decide to participate follow the game form to determine the levels of local public goods in the network formed by them.~\fn{This network is a subgraph obtained by removing the nodes corresponding to non-participating users from the original graph (directed network) constructed by all the users in the system.} In such a two stage mechanism, individual rationality implies the following. If the network formed by the participating users satisfies all the properties of Model~(M),~\fn{In particular, the network formed by the participating users must satisfy Assumption~\ref{chap5_Ass_G_indep_a} that there are at least three users affected by each local public good in this network. Note that all other assumptions of Model~(M) automatically carry over to the network formed by any subset of the users in Model~(M).} then, at all NE of the game induced by the game form among the participating users, the utility of each participating user will be at least as much as its utility without participation (i.e. if it is out of the system). 
    %This in turn imples that, if there are at least two other participating users that are affected by the actions of a user, then such a user voluntarily participates in the game form.       

    We would also like to clarify the rationale behind choosing NE as the solution concept for our problem. Note that because of assumptions \ref{chap5_Ass_Ai_pvt} and \ref{chap5_Ass_util_conc_pvt} in Model~(M), the environment of our problem is one of incomplete information. Therefore one may speculate the use of Bayesian Nash or dominant strategy as appropriate solution concepts for our problem. However, since the users in Model~(M) do not possess any prior beliefs about the utility functions and action sets of other users, we cannot use Bayesian Nash as a solution concept for Model~(M). Furthermore, because of impossibility results for the existence of non-parametric efficient dominant strategy mechanisms in classical public good environments \cite{Groves_Ledyard_1977}, we do not know if it is possible to design such mechanisms for the local public good environment of Model~(M). The well known Vickrey-Clarke-Groves (VCG) mechanisms that achieve incentive compatibility and efficiency with respect to non-numeraire goods, do not guarantee budget balance \cite{Groves_Ledyard_1977}. Hence they are inappropriate for our problem as budget balance is one of the desirable properties in our problem. VCG mechanisms are also unsuitable for our problem because they are direct mechanisms and any direct mechanism would require infinite message space to communicate the generic continuous (and concave) utility functions of users in Model~(M). Because of all of above reasons, and the known existence results for non-parametric, individually rational, budget-balanced Nash implementation mechanisms for classical private and public goods environments \cite{Groves_Ledyard_1977}, we choose Nash as the solution concept for our problem. 
    %%Because NE in general describe strategic behavior of users in games of complete information, we interpret NE in the incomplete information environment of Model~(M) in the way suggested by \cite[Section 4]{Groves_Ledyard_1977} and \cite{Reiter_Reichel_88}. Specifically, by quoting \cite[Section 4]{Groves_Ledyard_1977}, ``we do not suggest that each user knows all of system environment when it computes its message. We do suggest, however, that the complete information Nash game-theoretic equilibrium messages may be the possible stationary messages of some unspecified dynamic message exchange process.'' Alternatively,
    We adopt Nash's original ``mass action'' interpretation of NE \cite[page 21]{Nash_thesis}. Implicit in this interpretation is the assumption that the problems's environment is stable, that is, it does not change before the agents reach their equilibrium strategies. This assumption is consistent with our Assumption~\ref{chap5_Ass_net_const}. Nash's ``mass action'' interpretation of NE has also been adopted in \cite[pp. 69-70]{Groves_Ledyard_1977}, \cite[page 664]{Reiter_Reichel_88}, \cite{Sharma:2011}, and 
    \cite{Kakhbod_12a, Kakhbod_12b}.
    Specifically, by quoting \cite{Reiter_Reichel_88}, ``we interpret our analysis as applying to an unspecified (message exchange) process in which users grope their way to a stationary message and in which the Nash property is a necessary condition for stationarity.''

   We next construct a game form for the resource allocation problem ($P_D$) that achieves the abovementioned desirable properties -- Nash implementation, individual rationality, and budget balance.

%%%%%%%%%%%%%%%%%%%%%%%%%%%%%%%%%%%%%%%%%%%%%%%%%%%%%%%%%%%%%%%%%%%%%%%%%%%%%%%%%%%%%%%%%%%%%%%

\subsection{The game form} \label{chap5_Subsec_game_form}

    In this section we present a game form for the local public good provisioning problem presented in Section~\ref{chap5_Subsec_res_alloc_prb}. We provide explicit expressions of each of the components of the game form, the message space and the outcome function. We assume that the game form is common knowledge among the users and the network operator.

    {\bf The message space:} 
    Each user $i\in\N$ sends to the network operator a message $\Bm_i \in \Rl^{|\R_i|} \times \Rl_{+}^{|\R_i|} =: \M_i$ of the following form:
    \begin{eqnarray}
            \Bm_i := ( \tensor*[^{i}]{\ba}{_{\R_i}},  \tensor*[^{i}]{\bpi}{_{\R_i}} ); \quad
            \tensor*[^{i}]{\ba}{_{\R_i}} \in \Rl^{|\R_i|}, \;  \tensor*[^{i}]{\bpi}{_{\R_i}} \in \Rl_{+}^{|\R_i|},
        \label{chap5_Eq_mi}
    \end{eqnarray}
    \begin{equation}
        \mb{where,} \; \tensor*[^{i}]{\ba}{_{\R_i}} := ( \tensor*[^{i}]{a}{_k} )_{k \in \R_i}; \;
       % \mb{and} \ \
        \tensor*[^{i}]{\bpi}{_{\R_i}} := ( \tensor*[^{i}]{\pi}{_k} )_{k \in \R_i}, \; i \in \N.
        \label{chap5_Eq_i_p_Ai} 
    \end{equation}
    User $i$ also sends the component $(\tensor*[^{i}]{a}{_k}, \tensor*[^{i}]{\pi}{_k}), k \in \R_i,$ of its message to its neighbor $k \in \R_i$. In this message, $\tensor*[^{i}]{a}{_k}$ is the action proposal for user $k, k \in \R_i,$ by user $i, i \in \N$. Similarly, $\tensor*[^{i}]{\pi}{_k}$ is the price that user $i, i \in \N,$ proposes to pay for the action of user $k, k \in \R_i$. A detailed interpretation of these message elements is given in Section~\ref{chap5_Subsec_Thm}.  \\
    {\bf The outcome function:}
%
%
%    \begin{figure}[!ht]
%        \centering
%        \includegraphics[scale=0.30, page=2]{large_network.pdf}
%        \caption{Illustration of indexing rule for set $\C_j$ shown in Fig.~\ref{chap5_Fig_large_nw}. Index $\I_{rj}$ of user $r \in \C_j$ is indicated on the arrow directed from $j$ to $r$. The notation to denote these indices and to denote the user with a particular index is shown outside the dashed boundary demarcating the set $\C_j$.
%        }
%        \label{chap5_Fig_Cj_cyclic_index}
%    \end{figure}
%%{\bf Note: In this draft Fig. 2 appears on page 32.}
    After the users communicate their messages to the network operator, their actions and taxes are determined as follows. For each user $i \in \N$, the network operator determines the action $\h{a}_i$ of user $i$ from the messages communicated by its neighbors that are affected by it (set $\C_i$), i.e. from the message profile $\Bm_{\C_i} := (\Bm_k)_{k \in \C_i}$:
    \begin{equation}
        \h{a}_i(\Bm_{\C_i})  =  \frac{1}{|\C_i|}  \sum_{k \in \C_i}   \tensor*[^{k}]{a}{_i}, \quad i \in \N.
        \label{chap5_Eq_a_hat_i}
    \end{equation}

     To determine the users' taxes the network operator considers each set $\C_j$, $j \in \N$, and assigns indices $1, 2, \dots, |\C_j|$ in a cyclic order to the users in $\C_j$. Each index $1, 2, \dots, |\C_j|$ is assigned to an arbitrary but unique user $i \in \C_j$. Once the indices are assigned to the users in each set $\C_j$, they remain fixed throughout the time period of interest. We denote the index of user  $i$ associated with set $\C_j$ by $\I_{ij}$. The index $\I_{ij} \in \{ 1, 2, \dots, |\C_j| \}$ if $i \in \C_j$, and $\I_{ij} = 0$ if $i \nin \C_j$. Since for each set $\C_j,$ each index $1, 2, \dots, |\C_j|$ is assigned to a unique user $i \in \C_j$, therefore, $\fa i,k \in \C_j$ such that $i \neq k, \; \I_{ij} \neq \I_{kj}$. Note also that for any user $i \in \N$, and any $j, k \in \R_i$, the indices $\I_{ij}$ and $\I_{ik}$ are not necessarily the same and are independent of each other.  We denote the user with index $k \in \{ 1, 2, \dots, |\C_j| \}$ in set $\C_j$ by $\C_{j(k)}$. Thus, $\C_{j(\I_{ij})} = i$ for $i \in \C_j$. The cyclic order indexing means that, if $\I_{ij} = |\C_j|$, then  $\C_{j(\I_{ij}+1)} = \C_{j(1)}$, $\C_{j(\I_{ij}+2)} = \C_{j(2)}$, and so on. In Fig.~\ref{chap5_Fig_Cj_cyclic_index} we illustrate the above indexing rule for the set $\C_j$ shown in Fig.~\ref{chap5_Fig_large_nw}.

    Based on the above indexing, the users' taxes $\h{t}_i, i \in \N,$ are determined as follows. 
    \begin{equation}
    \begin{split}
         \h{t}_i( (\Bm_{\C_j})_{j \in \R_i} ) =  
       & \sum_{j \in \R_i} l_{ij}(\Bm_{\C_j})\; \h{a}_j(\Bm_{\C_j}) 
         + \sum_{j \in \R_i}  \tensor*[^{i}]{\pi}{_j}  \left( \tensor*[^{i}]{a}{_j} -   
         \tensor*[^{\C_{j(\I_{ij}+1)}}]{a}{_j} \right)^2  \\ 
       & - \sum_{j \in \R_i}  \tensor*[^{ \C_{j(\I_{ij}+1)} }]{\pi}{_j}  \left( \tensor*[^{\C_{j(\I_{ij}+1)}}]{a}{_j} - \tensor*[^{\C_{j(\I_{ij}+2)}}]{a}{_j} \right)^2  
          %i \in \N,
          \label{chap5_Eq_t_hat_i}
    \end{split}
    \end{equation}
    \begin{equation}
    \mb{ where,} \quad
       l_{ij}(\Bm_{\C_j})  =  \tensor*[^{ \C_{j(\I_{ij}+1)} }]{\pi}{_j}  -  \tensor*[^{ \C_{j(\I_{ij}+2)} }]{\pi}{_j},
       \ \ j \in \R_i, \ \ i \in \N.
       \label{chap5_Eq_l_ij} 
    \end{equation}

   We would like to emphasize here that the presence of the network operator is necessary for strategy-proofness and implementation of the above game form. A detailed discussion on the need and significance of the network operator is presented in Section~\ref{chap5_Subsec_impl_mech}.

    The game form given by (\ref{chap5_Eq_mi})--(\ref{chap5_Eq_l_ij}) and the users' aggregate utility functions in (\ref{chap5_Eq_u_i^A}) induce a game $(\times_{i\in\N}\M_i, (\h{a}_i, \h{t}_i)_{i\in\N} , \{u_i^A\}_{i \in \N})$. In this game, the set of network users $\N$ are the players, the set of strategies of a user is its message space $\M_i$, and a user's payoff is its utility $u_i^A\Big( \big( \h{a}_j( \Bm_{\C_j} ) \big)_{j \in \R_i},  \ \h{t}_i\big( (\Bm_{\C_j})_{j \in \R_i} \big) \Big)$ that it obtains at the allocation determined by the communicated messages. We define a NE of this game as a message profile $\Bm_{\N}^*$ that has the following property: $\fa i \in \N$ and $\fa \Bm_i \in \M_i,$
    \begin{equation}
    % \begin{split}
      \hs{-0.008in}   
        u_i^A\Big( \big( \h{a}_j( \Bm_{\C_j}^* ) \big)_{j \in \R_i}, \; 
        \h{t}_i\big( (\Bm_{\C_j}^*)_{j \in \R_i} \big) \Big)
           \geq 
         u_i^A\Big( \big( \h{a}_j( \Bm_i, \Bm_{\C_j}^* / i ) \big)_{j \in \R_i}, \;
               \h{t}_i\big( ( \Bm_i, \Bm_{\C_j}^* / i)_{j \in \R_i} \big) \Big). 
        \label{chap5_Eq_NE}
    % \end{split}
    \end{equation} 
    As discussed earlier, NE in general describe strategic behavior of users in games of complete information. This can be seen from (\ref{chap5_Eq_NE}) where, to define a NE, it requires complete information of all users' aggregate utility functions. However, the users in Model~(M) do not know each other's utilities; therefore, the game induced by the game form  (\ref{chap5_Eq_mi})--(\ref{chap5_Eq_l_ij}) and the users' aggregate utility functions (\ref{chap5_Eq_u_i^A}) is not one of complete information. Therefore, for our problem we adopt the NE interpretation of \cite{Reiter_Reichel_88} and \cite[Section 4]{Groves_Ledyard_1977} as discussed at the beginning of Section~\ref{chap5_Sec_dec_mech}. That is, we interpret NE as the ``stationary'' messages of an unspecified (message exchange) process that are characterized by the Nash property (\ref{chap5_Eq_NE}).

    In the next section we show that the allocations obtained by the game form presented in (\ref{chap5_Eq_mi})--(\ref{chap5_Eq_l_ij}) at all NE message profiles (satisfying (\ref{chap5_Eq_NE})), are optimal centralized allocations.

%%%%%%%%%%%%%%%%%%%%%%%%%%%%%%%%%%%%%%%%%%%%%%%%%%%%%%%%%%%%%%%%%%%%%%%%%%%%%%%%%%%%%%%%%%%%%%%%%%%%%%

\subsection{Properties of the game form} \label{chap5_Subsec_Thm}

    We begin this section with an intuitive discussion on how the game form presented in Section~\ref{chap5_Subsec_game_form} achieves optimal centralized allocations. We then formalize the results in Theorems~\ref{chap5_Thm_NE_opt} and \ref{chap5_Thm_NE_exists}.

    To understand how the proposed game form achieves optimal centralized allocations, let us look at the properties of NE allocations corresponding to this game form. A NE of the game induced by the game form (\ref{chap5_Eq_mi})--(\ref{chap5_Eq_l_ij}) and the users' utility functions (\ref{chap5_Eq_u_i^A}) can be interpreted as follows: Given the users' messages $\Bm_k, k \in \C_i$, the outcome function computes user $i$'s action as $1/|\C_i| \big( \sum_{k \in \C_i} \tensor*[^{k}]{a}{_i} \big)$. Therefore, user $i$'s action proposal $\tensor*[^{i}]{a}{_i}$ can be interpreted as the increment that $i$ desires over the sum of other users' action proposals for $i$, so as to bring its allocated action $\h{a}_i$ to its own desired value. Thus, if the action computed for $i$ based on its neighbors' proposals does not lie in $\A_i$, user $i$ can propose an appropriate action $\tensor*[^{i}]{a}{_i}$ and bring its allocated action within $\A_i$. The flexibility of proposing any action $\tensor*[^{i}]{a}{_i} \in \Rl$ gives each user $i \in \N$ the capability to bring its allocation $\h{a}_i$ within its feasible set $\A_i$ by unilateral deviation. Therefore, at any NE, $\h{a}_i \in \A_i, \fa i \in \N$. By taking the sum of taxes in (\ref{chap5_Eq_t_hat_i}) it can further be seen, after some computations, that the allocated tax profile $(\h{t}_i)_{i \in \N}$ satisfies (\ref{chap5_Eq_sum_ti_0}) (even at off-NE messages). Thus, all NE allocations $\Big( (\h{a}_i(\Bm_{\C_i}^*))_{i \in \N}, \; (\h{t}_i( (\Bm_{\C_j}^*)_{j \in \R_i} ))_{i \in \N}  \Big)$ lie in $\Do$ and hence are feasible solutions of Problem~($P_C$).

    To see further properties of NE allocations, let us look at the tax function in (\ref{chap5_Eq_t_hat_i}). The tax of user $i$ consists of three types of terms. The type-1 term is $\sum_{j \in \R_i} l_{ij}(\Bm_{\C_j})\; \h{a}_j(\Bm_{\C_j})$; it depends on all action proposals for each of user $i$'s neighbors $j \in \R_i$, and the price proposals for each of these neighbors by users other than user $i$. The type-2 term is $\sum_{j \in \R_i}  \tensor*[^{i}]{\pi}{_j}  \left( \tensor*[^{i}]{a}{_j} - \tensor*[^{\C_{j(\I_{ij}+1)}}]{a}{_j} \right)^2$; this term depends on $\tensor*[^{i}]{\ba}{_{\R_i}}$ as well as $\tensor*[^{i}]{\bpi}{_{\R_i}}$. Finally, the type-3 term is the following: $- \sum_{j \in \R_i}  \tensor*[^{ \C_{j(\I_{ij}+1)} }]{\pi}{_j} \times$ $\left( \tensor*[^{\C_{j(\I_{ij}+1)}}]{a}{_j} - \tensor*[^{\C_{j(\I_{ij}+2)}}]{a}{_j} \right)^2$; this term depends only on the messages of users other than $i$. Since $\tensor*[^{i}]{\bpi}{_{\R_i}}$ does not affect the determination of user $i$'s action, and affects only the type-2 term in $\h{t}_i$, the NE strategy of user $i, i \in \N$, that minimizes its tax is to propose for each $j \in \R_i$, $\tensor*[^{i}]{\bpi}{_j} = 0$ unless at the NE, $\tensor*[^{i}]{a}{_j} = \tensor*[^{\C_{j(\I_{ij}+1)}}]{a}{_j}$. Since the type-2 and type-3 terms in the users' tax are similar across users, for each $i \in \N$ and $j \in \R_i$, all the users $k \in \C_j$ choose the above strategy at NE. Therefore, the type-2 and type-3 terms vanish from every users' tax $\h{t}_i, i \in \N,$ at all NE. Thus, the tax that each user $i \in \N$ pays at a NE $\Bm_{\N}^*$ is of the form $\sum_{j \in \R_i} l_{ij}(\Bm_{\C_j}^*)\; \h{a}_j(\Bm_{\C_j}^*)$.  The NE term $l_{ij}(\Bm_{\C_j}^*), i \in \N, j \in \R_i$, can therefore be interpreted as the ``personalized price'' for user $i$ for the NE action $\h{a}_j(\Bm_{\C_j}^*)$ of its neighbor $j$. Note that at a NE, the personalized price for user $i$ is not controlled by $i$'s own message. The reduction of the users' NE taxes into the form $\sum_{j \in \R_i} l_{ij}(\Bm_{\C_j}^*)\; \h{a}_j(\Bm_{\C_j}^*)$ implies that at a NE, each user $i \in \N$ has a control over its aggregate utility only through its action proposal.~\fn{Note that user $i$'s action proposal determines the actions of all the users $j \in \R_i$; thus, it affects user $i$'s utility $u_i\Big( \big( \h{a}_j( \Bm_{\C_j}^* ) \big)_{j \in \R_i} \Big)$ as well as its tax $\sum_{j \in \R_i} l_{ij}(\Bm_{\C_j}^*)\; \h{a}_j(\Bm_{\C_j}^*)$.} If all other users' messages are fixed, each user has the capability of shifting the allocated action profile $\h{\ba}_{\R_i}$ to its desired value by proposing an appropriate $\tensor*[^{i}]{\ba}{_{\R_i}} \in \Rl^{|\R_i|}$ (See the discussion in the previous paragraph). Therefore, the NE strategy of each user $i \in \N$ is to propose an action profile $\tensor*[^{i}]{\ba}{_{\R_i}}$ that results in an allocation $\h{\ba}_{\R_i}$ that maximizes its aggregate utility.
        Thus, at a NE, each user maximizes its aggregate utility for its given personalized prices. By the construction of the tax function, the sum of the users' tax is zero at all NE and off equilibria. Thus, the individual aggregate utility maximization of the users also result in  the maximization of the sum of users' aggregate utilities subject to the budget balance constraint which is the objective of Problem~($P_C$).

    It is worth mentioning at this point the significance of type-2 and type-3 terms in the users' tax. As explained above, these two terms vanish at NE. However, if for some user $i \in \N$ these terms are not present in its tax $\h{t}_i$, then, the price proposal $\tensor*[^{i}]{\bpi}{_{\R_i}}$ of user $i$ will not affect its tax and hence, its aggregate utility. In such a case, user $i$ can propose arbitrary prices $\tensor*[^{i}]{\bpi}{_{\R_i}}$ because they would affect only other users' NE prices. The presence of type-2 and type-3 terms in user $i$'s tax prevent such a behavior as they impose a penalty on user $i$ if it proposes a high value of $\tensor*[^{i}]{\bpi}{_{\R_i}}$ or if its action proposal for its neighbors deviates too much from other users' proposals. Even though the presence of type-2 and type-3 terms in user $i$'s tax is necessary as explained above, it is important that the NE price $l_{ij}(\Bm_{\C_j}^*), j \in \R_i$ of user $i \in \N$ is not affected by $i$'s own proposal $\tensor*[^{i}]{\bpi}{_{\R_i}}$. This is because, in such a case, user $i$ may influence its own NE price in an unfair manner and may not behave as a price taker. To avoid such a situation, the type-2 and type-3 terms are designed in a way so that they vanish at NE. Thus, this construction induces price taking behavior in the users at NE and leads to optimal allocations.

    The results that formally establish the above properties of the game form are summarized in Theorems~\ref{chap5_Thm_NE_opt} and \ref{chap5_Thm_NE_exists} below.
    \begin{theorem} \label{chap5_Thm_NE_opt}
         Let $\Bm_{\N}^*$ be a NE of the game induced by the game form presented in Section~\ref{chap5_Subsec_game_form} and the users' utility functions (\ref{chap5_Eq_u_i^A}). Let $ (\h{\ba}_{\N}^*, \h{\bt}_{\N}^*) := ( \h{\ba}_{\N}(\Bm_{\N}^*), \h{\bt}_{\N}(\Bm_{\N}^*) )    := \Big( (\h{a}_i(\Bm_{\C_i}^*))_{i \in \N},$ $(\h{t}_i( (\Bm_{\C_j}^*)_{j \in \R_i} ))_{i \in \N}  \Big)$ be the action and tax profiles at $\Bm_{\N}^*$ determined by the game form. Then,
    \begin{itemize}
        \item[(a)] Each user $i \in \N$ weakly prefers its allocation $(\h{\ba}_{\R_i}^*, \h{t}_i^*)$ to the initial allocation $(\bm{0}, 0)$. Mathematically,
        \begin{equation*}
            \ u_i^A \Big( \h{\ba}_{\R_i}^*, \h{t}_i^* \Big)
            \geq u_i^A \Big( \bm{0},\; 0 \Big),
            \qquad \fa i \in \N.
            \label{chap5_Eq_Thm_NE_opt}
        \end{equation*}
        \item[(b)] $(\h{\ba}_{\N}^*, \h{\bt}_{\N}^*)$ is an optimal solution of Problem ($P_C$). \hfill $\boxtimes$ \vs{2mm}
    \end{itemize}
    \end{theorem} 
%%%%%%%%%%%%%%%%%%
    \begin{theorem} \label{chap5_Thm_NE_exists}
        Let $\h{\ba}_{\N}^*$ be an optimum action profile corresponding to Problem~($P_C$). Then,
    \begin{itemize}
        \item[(a)] There exist a set of personalized prices $l_{ij}^*, j\in\R_i, i\in\N,$ such that
            \begin{equation*}
                \h{\ba}_{\R_i}^* =
                \argmax_{ \stackrel{ \h{a}_i \in \A_i }{ \h{a}_j \in \Rl, \, j \in \R_i \bs \{i\} } }
                - \sum_{j \in \R_i} l_{ij}^*\; \h{a}_j + u_i(\h{\ba}_{\R_i}),
                \quad \fa i \in \N.
                \label{chap5_Eq_Thm_NE_exist_a}
           \end{equation*}

        \item[(b)] There exists at least one NE $\Bm_{\N}^*$ of the game induced by the game form presented in Section~\ref{chap5_Subsec_game_form} and the users' utility functions (\ref{chap5_Eq_u_i^A}) such that, $\h{\ba}_{\N}(\Bm_{\N}^*) = \h{\ba}_{\N}^*$. Furthermore, if $\h{t}_i^* := \sum_{j \in \R_i} l_{ij}^* \h{a}_j^*, i\in\N$, the set of all NE $\Bm_{\N}^* = (\Bm_i^*)_{i \in \N} = ( \tensor*[^{i}]{\ba}{_{\R_i}^*},  \tensor*[^{i}]{\bpi}{_{\R_i}^*} )$ that result in $(\h{\ba}_{\N}^*, \h{\bt}_{\N}^*)$ is characterized by the solution of the following set of conditions:
        \begin{eqnarray*}
            \frac{1}{|\C_i|}  \sum_{k \in \C_i}   \tensor*[^{k}]{a}{_i^*}
            & =& \h{a}_i^*, \quad i \in \N, \label{chap5_Thm2_NEchar_a_hat} \\
            {}^{ \C_{j(\I_{ij}+1)} }\pi_j^*  -  {}^{ \C_{j(\I_{ij}+2)} }\pi_j^*
            & =& l_{ij}^*,  \quad j \in \R_i, \ i \in \N, \label{chap5_Thm2_NEchar_lij} \\
            \tensor*[^{i}]{\pi}{_j^*}
            \left( \tensor*[^{i}]{a}{_j^*} - {}^{\C_{j(\I_{ij}+1)}}a_j^* \right)^2
            & =& 0, \quad j \in \R_i, \ \ i \in \N, \label{chap5_Thm2_NEchar_pi_a} \\
            \tensor*[^{i}]{\pi}{_j^*}
            & \geq& 0, \quad j \in \R_i, \ \ i \in \N. \hs{2in} \hfill \boxtimes \label{chap5_Thm2_NEchar_pi_pos}
        \end{eqnarray*}
    \end{itemize}
    \end{theorem}

    Because Theorem~\ref{chap5_Thm_NE_opt} is stated for an arbitrary NE $\Bm_{\N}^*$ of the game induced by the game form of Section~\ref{chap5_Subsec_game_form} and the users' utility functions (\ref{chap5_Eq_u_i^A}), the assertion of the theorem holds for all NE of this game. Thus, part~(a) of Theorem~\ref{chap5_Thm_NE_opt} establishes that the game form presented in Section~\ref{chap5_Subsec_game_form} is \emph{individually rational}, i.e., at any NE allocation, the aggregate utility of each user is at least as much as its aggregate utility before participating in the game/allocation process. Because of this property of the game form, each user voluntarily participates in the allocation process.
%%%%
%%%%
    Part~(b) of Theorem~\ref{chap5_Thm_NE_opt} asserts that all NE of the game induced by the game form of Section~\ref{chap5_Subsec_game_form} and the users' utility functions (\ref{chap5_Eq_u_i^A}) result in optimal centralized allocations (solutions of Problem~($P_C$)). Thus the set of NE allocations is a subset of the set of optimal centralized allocations. This establishes that the game form of Section~\ref{chap5_Subsec_game_form} \emph{implements in NE} the goal correspondence defined by the solutions of Problem~($P_C$). Because of this property, the above game form guarantees to provide an optimal centralized allocation irrespective of which NE is achieved in the game induced by it.

    The assertion of Theorem~\ref{chap5_Thm_NE_opt} that establishes the above two properties of the game form presented in Section~\ref{chap5_Subsec_game_form} is based on the assumption that there exists a NE of the game induced by this game form  and the users' utility functions (\ref{chap5_Eq_u_i^A}). However, Theorem~\ref{chap5_Thm_NE_opt} does not say anything about the existence of NE. Theorem~\ref{chap5_Thm_NE_exists} asserts that NE exist in the above game, and provides conditions that characterize the set of all NE that result in optimal centralized allocations of the form $(\h{\ba}_{\N}^*, \h{\bt}_{\N}^*) = (\h{\ba}_{\N}^*  , (\sum_{j \in \R_i} l_{ij}^* \h{a}_j^*)_{i \in \N} )$, where $\h{\ba}_{\N}^*$ is any optimal centralized action profile.
%%%%
%%%%
    In addition to the above, Theorem~\ref{chap5_Thm_NE_exists} also establishes the following property of the game form. Since the optimal action profile $\h{\ba}_{\N}^*$ in the statement of Theorem~\ref{chap5_Thm_NE_exists} is arbitrary, the theorem implies that the game form of Section~\ref{chap5_Subsec_game_form} can obtain each of the optimum action profiles of Problem~($P_C$) through at least one of the NE of the induced game. This establishes that the above game form is not biased towards any particular optimal centralized action profile.

We present the proofs of Theorem~\ref{chap5_Thm_NE_opt} and Theorem~\ref{chap5_Thm_NE_exists} in Appendices~\ref{chap5_Proof_Thm_NE_opt} and \ref{chap5_Proof_Thm_NE_exists}. In Appendix~\ref{Apx_illus_ex} we also present an example to illustrate how the properties established by Theorems~\ref{chap5_Thm_NE_opt} and \ref{chap5_Thm_NE_exists} are achieved by the proposed game form. 

%Due to lack of space, we are not presenting the proofs of Theorem~\ref{chap5_Thm_NE_opt} and Theorem~\ref{chap5_Thm_NE_exists} here. The proofs are available in \cite[Chapter 5]{Sharma_thesis}. 

    In the next section we present a discussion on how the proposed game form can be implemented in a real system, i.e., how the message communication and the determination of allocations specified by the game form can be carried out in a real system.

%%%%%%%%%%%%%%%%%%%%%%%%%%%%%%%%%%%%%%%%%%%%%%%%%%%%%%%%%%%%
\subsection{Implementation of the game form} \label{chap5_Subsec_impl_mech}
%%%%%%%%%%%%%%%%%%%%%%%%%%%%%%%%%%%%%%%%%%%%%%%%%%%%%%%%%%%%

%    In this section we discuss two aspects of implementation of the decentralized mechanism specified by the game form of Section~\ref{chap5_Subsec_game_form}. First we discuss how the game form itself can be implemented,  We then discuss how NE can be achieved in the game induced by the above game form.

    We explain in this section how the game form of Section~\ref{chap5_Subsec_game_form} can be implemented in the presence of a network operator, and why the presence of a network operator is necessary for the implementation of the game form. To this end let us first suppose that the network operator is not present in the network. As discussed in Section~\ref{chap5_Subsec_game_form} the outcome function specifies the allocation $(\h{a}_i, \h{t}_i)$ for a user $i \in \N$ based on its neighbors' messages. Since the game form is common knowledge among the users, if each user announces its messages to all its neighbors, every user can have the required set of messages to compute its own allocations. However, with this kind of local communication, the messages required to compute user $i$'s allocation are not necessarily known to users other than $i$. Therefore, even though the other users know the outcome function for user $i$, no other user can check if the allocation determined by user $i$ corresponds to its neighbors' messages. Since each user $i \in \N$ is selfish, it cannot be relied upon for the determination of its allocation. Therefore, in large-scale systems such as one represented by Model~(M), where each user does not hear all other users' messages, the presence of a network operator is extremely important. The network operator's role is twofold. First, according to the specification of the game form (of Section~\ref{chap5_Subsec_game_form}) each user announces its messages to its neighbors as well as to the network operator. The network operator knows the network structure (Assumption~\ref{chap5_Ass_nbr_know}) and the outcome function for each user. Thus, it can compute all the allocations based on the messages it receives, and then it can tell each user its corresponding allocation (or it can check whether the allocation $(a_i^*, t_i^*)$ implemented by user $i, i \in \N$, is the same as that specified by the mechanism).
		The other role of the network operator that facilitates implementation of the game form is the following. Note that the game form specifies redistribution of money among the users by charging each user an appropriate positive or negative tax (see (\ref{chap5_Eq_sum_ti_0})). This means that the tax money must go from one subset of the users to the other subset of users. Since the users do not have complete network information, nor do they know the allocations of other users in the network, they cannot determine the appropriate flow of money in the network. The network operator implements this redistribution of money by acting as an accountant that collects money from the users who must pay positive tax according to the game form and gives the money back to the users who must receive subsidies (negative tax).

%%%%%%%%%%%%%%%%%%%%%%%%%%%%%%%%%%%%%%%%%%%%%%%%%%%%%%%%%%%%%%%%%%%%%%%%
\section{Future directions} \label{chap5_Sec_conc} 
%%%%%%%%%%%%%%%%%%%%%%%%%%%%%%%%%%%%%%%%%%%%%%%%%%%%%%%%%%%%%%%%%%%%%%%%

The problem formulation and the solution of the local public goods provisioning problem presented in this paper open up several new directions for future research. First, the development of efficient mechanisms that can compute NE is an important open problem. To address this problem there can be two different directions. (i) The development of algorithms that guarantee convergence to Nash equilibria of the games constructed in this paper. (ii)  The development of alternative mechanisms/game forms that lead to games with dynamically stable NE. Second, the network model we studied in this paper assumed a given set of users and a given network topology. In many local public good networks such as social or research networks, the set of network users and the network topology must be determined as part of network objective maximization. These situations give rise to interesting admission control and network formation problems many of which are open research problems.  Finally, in this paper we focused on static resource allocation problem where the characteristics of the local public good network do not change with time. The development of implementation mechanisms under dynamic situations, where the network characteristics change during the determination of resource allocation, are open research problems. \\
%\mb{} 
%%%%%%%%%%%%%%%%%%%%%%%%%%%%%%%%%%%%%%%%%%%%%%%%%%%%%%%%%%%%%%%%%%%%%%%%
{\bf Acknowledgments:} \hs{1mm} 
This work was supported in part by NSF grant CCF-1111061. The authors are grateful to Y. Chen and A. Anastasopoulos at the University of Michigan for stimulating discussions.

\bibliography{Shruti_full}

%%%%%%%%%%%%%%%%%%%%%%%%%%%%%%%%%%%%%%%%%%%%%%%%%%%%%%%%%%%%
\appendices
%%%%%%%%%%%%%%%%%%%%%%%%%%%%%%%%%%%%%%%%%%%%%%%%%%%%%%%%%%%%

    In Appendices~\ref{chap5_Proof_Thm_NE_opt} and \ref{chap5_Proof_Thm_NE_exists} that follow, we present the proof of Theorems~\ref{chap5_Thm_NE_opt} and \ref{chap5_Thm_NE_exists}, respectively. We divide the proof into several claims to organize the presentation. In Appendix~\ref{Apx_illus_ex} we present an example to illustrate how the properties established by Theorems~\ref{chap5_Thm_NE_opt} and \ref{chap5_Thm_NE_exists} are achieved by the proposed game form. 
%% In the paper put reference to this archive for applications and example.     

%%%%%%%%%%%%%%%%%%%%%%%%%%%%%%%%%%%%%%%%%%%%%%%%%%%%%%%%%%%%
\section{Proof of Theorem~1} \label{chap5_Proof_Thm_NE_opt}
%%%%%%%%%%%%%%%%%%%%%%%%%%%%%%%%%%%%%%%%%%%%%%%%%%%%%%%%%%%%

    We prove Theorem~\ref{chap5_Thm_NE_opt} in four claims. In Claims~\ref{chap5_Cl_NE_tax_form} and \ref{chap5_Cl_indv_ratn} we show that all users weakly prefer a NE allocation (corresponding to the game form presented in Section~\ref{chap5_Subsec_game_form}) to their initial allocations; these claims prove part~(a) of Theorem~\ref{chap5_Thm_NE_opt}. In Claim~\ref{chap5_Cl_fes_sol} we show that a NE allocation is a feasible solution of Problem~($P_C$). In Claim~\ref{chap5_Cl_NE_optl} we show that a NE action profile is an optimal action profile for Problem~($P_C$). Thus, Claim~\ref{chap5_Cl_fes_sol} and Claim~\ref{chap5_Cl_NE_optl} establish that a NE allocation is an optimal solution of Problem~($P_C$) and prove part~(b) of Theorem~\ref{chap5_Thm_NE_opt}.

%%%%%%%%%%%%%%%%%%%%%%%%%%%%%%%%%%%%%%%%%%%%%%%%%%%%%%%%%%%%
\begin{Cl} \label{chap5_Cl_fes_sol}
		If $\Bm_{\N}^{*}$  is a NE of the game induced by the game form presented in Section~\ref{chap5_Subsec_game_form} and the users' utility functions (\ref{chap5_Eq_u_i^A}), then the action and tax profile $(\h{\ba}_{\N}^*, \h{\bt}_{\N}^*) := (\h{\ba}_{\N}(\Bm_{\N}^*), \h{\bt}_{\N}(\Bm_{\N}^*) )$ is a feasible solution of Problem~($P_C$), i.e. $(\h{\ba}_{\N}^*, \h{\bt}_{\N}^*) \in \Do$.
\end{Cl}
%%%%%%%%%%%%%%%%%%%%%%%%%%%%%%%%%%%%%%%%%%%%%%%%%%%%%%%%%%%%
{\bf Proof:} 
We prove the feasibility of the NE action and tax profiles in two steps. First we prove the feasibility of the NE tax profile, then we prove the feasibility of the NE action profile.

To prove the feasibility of NE tax profile, we need to show that it satisfies (\ref{chap5_Eq_sum_ti_0}). For this, we first take the sum of second and third terms on the Right Hand Side (RHS) of (\ref{chap5_Eq_t_hat_i}) over all $i \in \N$, i.e.
\begin{equation}
\begin{split}
		\sum_{i \in \N}  \sum_{j \in \R_i} \Bigg[
    \tensor*[^{i}]{\pi}{_j}  \left(\tensor*[^{i}]{a}{_j} - 	
    \tensor*[^{\C_{j(\I_{ij}+1)}}]{a}{_j}\right)^2 
   -\tensor*[^{ \C_{j(\I_{ij}+1)} }]{\pi}{_j}  \left( 	
    \tensor*[^{\C_{j(\I_{ij}+1)}}]{a}{_j} -
    \tensor*[^{\C_{j(\I_{ij}+2)}}]{a}{_j} \right)^2  \Bigg].
    \label{chap5_Eq_sum_ti_23_one}
\end{split}
\end{equation}
From the construction of the graph matrix $\G$ and the sets $\R_i$ and $\C_j$, $i,j \in \N$, the sum $\sum_{i \in \N}  \sum_{j \in \R_i} (\cdot)$  is equal to the sum $\sum_{j \in \N}  \sum_{i \in \C_j} (\cdot)$. Therefore, we can rewrite (\ref{chap5_Eq_sum_ti_23_one}) as
\begin{equation}
\begin{split}
    \sum_{j \in \N}  \Bigg[
    \sum_{i \in \C_j} \tensor*[^{i}]{\pi}{_j}  \left(\tensor*[^{i}]{a}{_j} - 	
    \tensor*[^{\C_{j(\I_{ij}+1)}}]{a}{_j}\right)^2
  - \sum_{i \in \C_j} \tensor*[^{ \C_{j(\I_{ij}+1)} }]{\pi}{_j}  \left( 
    \tensor*[^{\C_{j(\I_{ij}+1)}}]{a}{_j} -
    \tensor*[^{\C_{j(\I_{ij}+2)}}]{a}{_j} \right)^2  \Bigg].
    \label{chap5_Eq_sum_ti_23_two}
\end{split}
\end{equation}
Note that both the sums inside the square brackets in (\ref{chap5_Eq_sum_ti_23_two}) are over all $i \in \C_j$. Because of the cyclic indexing of the users in each set $\C_j$, $j \in \N$, these two sums are equal. Therefore the overall sum in (\ref{chap5_Eq_sum_ti_23_two}) evaluates to zero. Thus, the sum of taxes in (\ref{chap5_Eq_t_hat_i}) reduces to
\begin{equation}
\begin{split}
    \sum_{i \in \N}  \h{t}_i( (\Bm_{\C_j})_{j \in \R_i})  =
    \sum_{i \in \N} \sum_{j \in \R_i} l_{ij}(\Bm_{\C_j})\; \h{a}_j(\Bm_{\C_j}).
    \label{chap5_Eq_sum_ti_1_one}
\end{split}
\end{equation}
Combining (\ref{chap5_Eq_l_ij}) and (\ref{chap5_Eq_sum_ti_1_one}) we obtain
\begin{equation}
\begin{split}
     \sum_{i \in \N}  \h{t}_i( (\Bm_{\C_j})_{j \in \R_i})
   = \sum_{j \in \N} \Bigg[ \sum_{i \in \C_j} \tensor*[^{ \C_{j(\I_{ij}+1)} }]{\pi}{_j} 
   - \sum_{i \in \C_j} \tensor*[^{ \C_{j(\I_{ij}+2)} }]{\pi}{_j}  \Bigg] \h{a}_j(\Bm_{\C_j})
   = 0.
     \label{chap5_Eq_sum_ti_1_two}
\end{split}
\end{equation}
The second equality in (\ref{chap5_Eq_sum_ti_1_two}) follows because of the cyclic indexing of the users in each set $\C_j$, $j \in \N$, which makes the two sums inside the square brackets in (\ref{chap5_Eq_sum_ti_1_two}) equal. Because (\ref{chap5_Eq_sum_ti_1_two}) holds for any arbitrary message profile $\Bm_{\N}$, it follows that at NE $\Bm_{\N}^*$,
\begin{equation}
    \sum_{i \in \N}  \h{t}_i( (\Bm_{\C_j}^*)_{j \in \R_i}) = 0.
    \label{chap5_Eq_sum_ti_NE_0}
\end{equation}

To complete the proof of Claim~\ref{chap5_Cl_fes_sol}, we have to prove that for all $i \in \N$, $\h{a}_i(\Bm_{\C_i}^*) \in \A_i$. We prove this by contradiction.
Suppose $\h{a}_i^* \nin \A_i$ for some $i \in \N$. Then, from (\ref{chap5_Eq_u_i^A}), $u_i^A( \h{\ba}_{\R_i}^*, \h{t}_i^* ) = -\infty$. Consider $\wt{\Bm}_i = ( ( \tensor*[^{i}]{\wt{a}}{_i},  \tensor*[^{i}]{\ba}{_{\R_i}^*} / i ),  \tensor*[^{i}]{\bpi}{_{\R_i}^*} )$ where $\tensor*[^{i}]{a}{_k^*}, k \in \R_i \bs \{i\},$ and $\tensor*[^{i}]{\bpi}{_{\R_i}^*}$ are respectively  the NE action and price proposals of user $i$ and $\tensor*[^{i}]{\wt{a}}{_i}$ is such that
\begin{equation}
     \h{a}_i(\wt{\Bm}_i,  \Bm_{\C_i}^*/i)  
   = \frac{1}{|\C_i|}  \Big( \tensor*[^{i}]{\wt{\ba}}{_i} +
     \sum_{\stackrel{k\in\C_i}{ k\neq i}}  \tensor*[^{k}]{a}{_i^*} \Big)
     \in  \A_i.
     \label{chap5_Eq_Cl1_a_m_tilda}
\end{equation}
Note that the flexibility of user $i$ in choosing any message $\tensor*[^{i}]{\ba}{_{\R_i}} \in \Rl^{|\R_i|}$ (see (\ref{chap5_Eq_mi})) allows it to choose an appropriate  $\tensor*[^{i}]{\wt{a}}{_i}$ that satisfies the condition in (\ref{chap5_Eq_Cl1_a_m_tilda}). For the message $\wt{\Bm}_i$ constructed above,
\begin{equation}
\begin{split}
      & u_i^A \Big( \big(\h{a}_{k}( \wt{\Bm}_i,  \Bm_{\C_k}^*/i ) \big)_{k\in\R_i},\;\;
        \h{t}_i \big(( \wt{\Bm}_i , \,\Bm_{\C_j}^* / i )_{j \in \R_i} \big) \Big)  \\
    = & - \h{t}_i \big(( \wt{\Bm}_i , \,\Bm_{\C_j}^* / i )_{j \in \R_i} \big)
        + u_i \Big( \big(\h{a}_{k}( \wt{\Bm}_i,  \Bm_{\C_k}^*/i ) \big)_{k\in\R_i} \Big)
    >  - \infty = u_i^A( \h{\ba}_{\R_i}^*, \h{t}_i^* ).
    \label{chap5_Eq_Cl1_user_dev}
\end{split}
\end{equation}
Thus if $\h{a}_i(\Bm_{\C_i}^*) \nin \A_i$ user $i$ finds it profitable to deviate to $\wt{\Bm}_{\bi}$. Inequality~(\ref{chap5_Eq_Cl1_user_dev}) implies that $\Bm_{\N}^*$ cannot be a NE, which is a contradiction. Therefore, at any NE $\Bm_{\N}^*$, we must have $\h{a}_i(\Bm_{\C_i}^*) \in \A_i \fa i \in \N$. This along with (\ref{chap5_Eq_sum_ti_NE_0}) implies that, $(\h{\ba}_{\N}^*, \h{\bt}_{\N}^*) \in \Do$.
\hfill \fbox{}

%%%%%%%%%%%%%%%%%%%%%%%%%%%%%%%%%%%%%%%%%%%%%%%%%%%%%%%%%%%%
\begin{Cl} \label{chap5_Cl_NE_tax_form}
    If $\Bm_{\N}^{*}$  is a NE of the game induced by the game form presented in Section~\ref{chap5_Subsec_game_form} and the users' utility functions (\ref{chap5_Eq_u_i^A}), then, the tax $\h{t}_i( (\Bm_{\C_j}^*)_{j \in \R_i}) =: \h{t}_i^*$ paid by user $i, i \in \N,$ at the NE $\Bm_{\N}^*$ is of the form $\h{t}_i^* = \sum_{j \in \R_i} l_{ij}^*\; \h{a}_j^*$, where $l_{ij}^* = l_{ij}(\Bm_{\C_j}^*)$ and $\h{a}_j^* = \h{a}_j(\Bm_{\C_j}^*)$.
\end{Cl}
%%%%%%%%%%%%%%%%%%%%%%%%%%%%%%%%%%%%%%%%%%%%%%%%%%%%%%%%%%%%
{\bf Proof:}
Let $\Bm_{\N}^*$ be the NE specified in the statement of Claim~\ref{chap5_Cl_NE_tax_form}. Then, for each $i\in\N$,
\begin{equation}
 \begin{split}
     u_i^A \Big( \big(\h{a}_{k}( \Bm_i,  \Bm_{\C_k}^*/i )  \big)_{k\in\R_i}, \;
     \h{t}_i \big(( \Bm_i ,  \,\Bm_{\C_j}^* / i )_{j \in \R_i} \big) \Big) 
     \leq \ u_i^A \Big( \h{\ba}_{\R_i}^*, \h{t}_i^* \Big), \, \fa \Bm_i \in \M_i.
     \label{chap5_Cl2_uiA_NE_dev}
 \end{split}
\end{equation}
Substituting $\Bm_i = ( \tensor*[^{i}]{\ba}{_{\R_i}^*} , \tensor*[^{i}]{\bpi}{_{\R_i}} ), \ \tensor*[^{i}]{\bpi}{_{\R_i}} \in \Rl_+^{|\R_i|}$, in (\ref{chap5_Cl2_uiA_NE_dev}) and using (\ref{chap5_Eq_a_hat_i}) implies
that
\begin{equation}
 \begin{split}
     u_i^A \Big( \h{\ba}_{\R_i}^*, \;\; \h{t}_i \big( ( (\tensor*[^{i}]{\ba}{_{\R_i}^*},  
     \tensor*[^{i}]{\bpi}{_{\R_i}}),  \, \Bm_{\C_j}^* / i )_{j \in \R_i} \big) \Big) 
     \leq \ u_i^A \Big( \h{\ba}_{\R_i}^*, \h{t}_i^* \Big), \quad
     \fa \tensor*[^{i}]{\bpi}{_{\R_i}} \in \Rl_+^{|\R_i|}.
 \label{chap5_Cl2_uiA_NE_pi_dev}
 \end{split}
\end{equation}
Since $u_i^A$ decreases in $t_i$ (see (\ref{chap5_Eq_u_i^A})), (\ref{chap5_Cl2_uiA_NE_pi_dev}) implies that
\begin{equation}
 \begin{split}
    \h{t}_i \Big( \big( (\tensor*[^{i}]{\ba}{_{\R_i}^*} , \tensor*[^{i}]{\bpi}{_{\R_i}}), \, \Bm_{\C_j}^* / i \big)_{j \in \R_i} \Big)
        \geq  \h{t}_i^*, \quad
        \fa \tensor*[^{i}]{\bpi}{_{\R_i}} \in \Rl_+^{|\R_i|}.
 \label{chap5_Cl2_ti_NE_pi_dev}
 \end{split}
\end{equation}
Substituting (\ref{chap5_Eq_t_hat_i}) in (\ref{chap5_Cl2_ti_NE_pi_dev}) results in
\begin{equation}
 \begin{split}
       \sum_{j \in \R_i} \bigg[ l_{ij}^* \h{a}_j^* 
       + \tensor*[^{i}]{\pi}{_j}
       \left( \tensor*[^{i}]{a}{_j^*} - {}^{\C_{j(\I_{ij}+1)}}a_j^* \right)^2 
       - {}^{ \C_{j(\I_{ij}+1)} }\pi_j^*
       \left( {}^{\C_{j(\I_{ij}+1)}}a_j^* - {}^{\C_{j(\I_{ij}+2)}}a_j^* \right)^2 \bigg] & \geq \\        
       \sum_{j \in \R_i} \bigg[ l_{ij}^* \h{a}_j^*
       + \tensor*[^{i}]{\pi}{_j^*}
       \left( \tensor*[^{i}]{a}{_j^*} - {}^{\C_{j(\I_{ij}+1)}}a_j^* \right)^2
       - {}^{ \C_{j(\I_{ij}+1)} }\pi_j^*
       \left( {}^{\C_{j(\I_{ij}+1)}}a_j^* - {}^{\C_{j(\I_{ij}+2)}}a_j^* \right)^2 \bigg] &, 
       \fa \tensor*[^{i}]{\bpi}{_{\R_i}} \in \Rl_+^{|\R_i|}. 
 \label{chap5_Cl2_ti_exp_NE}
 \end{split}
\end{equation}
Canceling the common terms in (\ref{chap5_Cl2_ti_exp_NE}) gives
\begin{equation}
 \begin{split}
    \sum_{j \in \R_i}
    ( \tensor*[^{i}]{\pi}{_j} - \tensor*[^{i}]{\pi}{_j^*} )
    \left( \tensor*[^{i}]{a}{_j^*} - {}^{\C_{j(\I_{ij}+1)}}a_j^* \right)^2
    \geq 0, \qquad
    \fa \tensor*[^{i}]{\bpi}{_{\R_i}} \in \Rl_+^{|\R_i|}.
 \label{chap5_Cl2_ti_cancel_NE}
 \end{split}
\end{equation}
Since (\ref{chap5_Cl2_ti_cancel_NE}) must hold for all $\tensor*[^{i}]{\bpi}{_{\R_i}} \in \Rl_+^{|\R_i|}$, we must have that
\begin{equation}
 \begin{split}
    \mb{for each} \ j \in \R_i, \ \mb{either} \
    \tensor*[^{i}]{\pi}{_j^*} = 0 \ \mb{or} \
        \tensor*[^{i}]{a}{_j^*} = {}^{\C_{j(\I_{ij}+1)}}a_j^*.
 \label{chap5_Cl2_pi_a_0}
 \end{split}
\end{equation}
From (\ref{chap5_Cl2_pi_a_0}) it follows that at any NE $\Bm_{\N}^*$,
\begin{equation}
 \begin{split}
    \tensor*[^{i}]{\pi}{_j^*}
        \left( \tensor*[^{i}]{a}{_j^*} - {}^{\C_{j(\I_{ij}+1)}}a_j^* \right)^2
        = 0,
    \ \ \fa j \in \R_i, \ \ \fa i \in \N.
 \label{chap5_Cl2_pi_a_0_NE}
 \end{split}
\end{equation}
Note that (\ref{chap5_Cl2_pi_a_0_NE}) also implies that $\fa i \in \N$ and $\fa j \in \R_i$,
\begin{equation}
 \begin{split}
    {}^{\C_{j(\I_{ij}+1)}}\pi_j^*
        \left( {}^{\C_{j(\I_{ij}+1)}}a_j^* - {}^{\C_{j(\I_{ij}+2)}}a_j^* \right)^2
        = 0.
 \label{chap5_Cl2_pi_a_next_0_NE}
 \end{split}
\end{equation}
(\ref{chap5_Cl2_pi_a_next_0_NE}) follows from (\ref{chap5_Cl2_pi_a_0_NE}) because for each $i \in \N$,  $j \in \R_i$ also implies that $j \in \R_{\C_{j(\I_{ij}+1)}}$.
Using (\ref{chap5_Cl2_pi_a_0_NE}) and (\ref{chap5_Cl2_pi_a_next_0_NE}) in (\ref{chap5_Eq_t_hat_i}) we obtain that any NE tax profile must be of the form
\begin{equation}
    \h{t}_i^* = \sum_{j \in \R_i} l_{ij}^*\; \h{a}_j^*, \quad \fa i \in \N.
\label{chap5_Cl2_NE_tax_form}
\end{equation}
\hfill \fbox{}

%%%%%%%%%%%%%%%%%%%%%%%%%%%%%%%%%%%%%%%%%%%%%%%%%%%%%%%%%%%%
\begin{Cl} \label{chap5_Cl_indv_ratn}
    The game form given in Section~\ref{chap5_Subsec_game_form} is individually rational, i.e. at every NE $\Bm_{\N}^*$ of the game induced by this game form and the users' utilities in (\ref{chap5_Eq_u_i^A}), each user $i \in \N$ weakly prefers the allocation $(\h{\ba}_{\R_i}^*, \h{t}_i^*)$ to the initial allocation $(\bm{0},0)$. Mathematically,
    \begin{equation}
        u_i^A \Big( \bm{0},\; 0 \Big)
        \leq \ u_i^A \Big( \h{\ba}_{\R_i}^*, \h{t}_i^* \Big),
        \qquad \fa i \in \N.
    \end{equation}
\end{Cl}
%%%%%%%%%%%%%%%%%%%%%%%%%%%%%%%%%%%%%%%%%%%%%%%%%%%%%%%%%%%%
{\bf Proof:}
Suppose $\Bm_{\N}^{*}$  is a NE of the game induced by the game form presented in Section~\ref{chap5_Subsec_game_form} and the users' utility functions (\ref{chap5_Eq_u_i^A}). From Claim~\ref{chap5_Cl_NE_tax_form} we know the form of users' tax at $\Bm_{\N}^*$. Substituting that from (\ref{chap5_Cl2_NE_tax_form}) into (\ref{chap5_Cl2_uiA_NE_dev}) we obtain that for each $i \in \N$,
\begin{equation}
 \begin{split}
        u_i^A \Big( \big(\h{a}_{k}( \Bm_i,  \Bm_{\C_k}^*/i ) \big)_{k\in\R_i},\;\;
        \h{t}_i \big(( \Bm_i , \,\Bm_{\C_j}^* / i )_{j \in \R_i} \big) \Big) 
   \leq u_i^A \Big( \h{\ba}_{\R_i}^*, \sum_{j \in \R_i} l_{ij}^*\; \h{a}_j^* \Big) &, \\ 
   \fa \Bm_i = ( \tensor*[^{i}]{\ba}{_{\R_i}},  \tensor*[^{i}]{\bpi}{_{\R_i}} ) \in \M_i &.
 \label{chap5_Cl3_uiA_NE_dev}
 \end{split}
\end{equation}
Substituting for $\h{t}_i$ in (\ref{chap5_Cl3_uiA_NE_dev}) from (\ref{chap5_Eq_t_hat_i}) and using (\ref{chap5_Cl2_pi_a_next_0_NE}) we obtain,
\begin{equation}
 \begin{split}
   & u_i^A \bigg( 
      \Big(\h{a}_{k} \big( ( \tensor*[^{i}]{\ba}{_{\R_i}}, 
      \tensor*[^{i}]{\bpi}{_{\R_i}} ), \Bm_{\C_k}^*/i \big) \Big)_{k\in\R_i},\;\;  
      \sum_{j \in \R_i} \! \Big( l_{ij}^*\; \h{a}_j \big(( \tensor*[^{i}]{\ba}{_{\R_i}},
      \tensor*[^{i}]{\bpi}{_{\R_i}}), \Bm_{\C_j}^*/i  \big) 
      + \tensor*[^{i}]{\pi}{_j}  \big( \tensor*[^{i}]{a}{_j}  
      - \tensor*[^{\C_{j(\I_{ij}+1)}}]{a}{_j} \big)^2  \Big) \! \bigg) \\
   &   \leq u_i^A \Big( 
     \h{\ba}_{\R_i}^*, \sum_{j \in \R_i} l_{ij}^*\; \h{a}_j^* \Big),
      \qquad \qquad \qquad \qquad \qquad \qquad 
      \fa \tensor*[^{i}]{\ba}{_{\R_i}} \in \Rl^{|\R_i|},
      \fa \tensor*[^{i}]{\bpi}{_{\R_i}} \in \Rl_+^{|\R_i|}.
 \label{chap5_Cl3_uiA_ti_exp_dev}
 \end{split}
\end{equation}
In particular, $\tensor*[^{i}]{\bpi}{_{\R_i}} = \bm{0}$ in (\ref{chap5_Cl3_uiA_ti_exp_dev}) implies that
\begin{equation}
 \begin{split}
         u_i^A \bigg( \Big(\h{a}_{k} \big( ( \tensor*[^{i}]{\ba}{_{\R_i}}, \bm{0} ),
          \Bm_{\C_k}^*/i \big) \Big)_{k\in\R_i}, \; 
          \sum_{j \in \R_i} \Big( l_{ij}^*\; \h{a}_j \big(
          ( \tensor*[^{i}]{\ba}{_{\R_i}}, \bm{0} ), \Bm_{\C_j}^*/i  \big) \Big) \bigg) 
 \leq \    u_i^A \Big( \h{\ba}_{\R_i}^*, \sum_{j \in \R_i} l_{ij}^*\; \h{a}_j^* \Big),
      %\qquad \qquad \qquad \qquad \qquad \qquad \qquad  
          \fa \tensor*[^{i}]{\ba}{_{\R_i}} \in \Rl^{|\R_i|}.
 \label{chap5_Cl3_uiA_pi_0_dev}
 \end{split}
\end{equation}
Since (\ref{chap5_Cl3_uiA_pi_0_dev}) holds for all $\tensor*[^{i}]{\ba}{_{\R_i}} \in \Rl^{|\R_i|}$, substituting in it $\frac{1}{|\C_j|}  ( \tensor*[^{i}]{a}{_j} + \sum_{k \in \C_j \bs \{i\} }   \tensor*[^{k}]{a}{_j^*} ) = \ol{a}_j \fa j \in \R_i$ implies, 
\begin{equation}
 \begin{split}
    u_i^A \Big( \big( \ol{a}_j \big)_{j \in \R_i},\;
        \sum_{j \in \R_i} \big( l_{ij}^*\; \ol{a}_j \big) \Big)
        \leq \ u_i^A \Big( \h{\ba}_{\R_i}^*, \sum_{j \in \R_i} l_{ij}^*\; \h{a}_j^* \Big), 
        \; \fa \ol{a}_{\R_i} := (\ol{a}_j)_{j \in \R_i} \in \Rl^{|\R_i|}. 
 \label{chap5_Cl3_iaj_subs_dev}
 \end{split}
\end{equation}
For $\ol{a}_{\R_i} = \bm{0}$, (\ref{chap5_Cl3_iaj_subs_dev}) implies further that
\begin{equation}
 \begin{split}
    u_i^A \Big( \bm{0},\; 0 \Big)
        \leq \ u_i^A \Big( \h{\ba}_{\R_i}^*, \sum_{j \in \R_i} l_{ij}^*\; \h{a}_j^* \Big),
        \qquad \qquad\qquad\qquad\qquad \fa i \in \N.
 \label{chap5_Cl3_ind_ratn}
 \end{split}
\end{equation}
\hfill \fbox{}

%%%%%%%%%%%%%%%%%%%%%%%%%%%%%%%%%%%%%%%%%%%%%%%%%%%%%%%%%%%%
\begin{Cl} \label{chap5_Cl_NE_optl}
   A NE allocation $(\h{\ba}_{\N}^*, \h{\bt}_{\N}^*)$ is an optimal solution of the centralized problem~$(P_{C})$.
\end{Cl}
%%%%%%%%%%%%%%%%%%%%%%%%%%%%%%%%%%%%%%%%%%%%%%%%%%%%%%%%%%%%
{\bf Proof:}
For each $i \in \N$, (\ref{chap5_Cl3_iaj_subs_dev}) can be equivalently written as
\begin{equation}
 \begin{split}
    \h{\ba}_{\R_i}^*
    & \in \argmax_{\ol{\ba}_{\R_i} \in \Rl^{|\R_i|}}
        u_i^A \Big( \ol{\ba}_{\R_i},\; \sum_{j \in \R_i} l_{ij}^*\; \ol{a}_j \Big) \\
    & = \argmax_{ \stackrel{ \ol{a}_i \in \A_i }{ \ol{a}_j \in \Rl, \, j \in \R_i \bs \{i\} } }
        \left\{ - \sum_{j \in \R_i} l_{ij}^*\; \ol{a}_j + u_i(\ol{\ba}_{\R_i}) \right\}
 \label{chap5_Cl4_aRi*_argmax}
 \end{split}
\end{equation} 
Let for each $i \in \N,$ $f_{\A_i}(a_i)$ be a convex function that characterizes the set $\A_i$ as, $a_i \in \A_i \Leftrightarrow f_{\A_i}(a_i) \leq 0$.~\fn{By \cite{Boyd} we can always find a convex function that characterizes a convex set.}

Since for each $i \in \N$, $u_i( \ol{\ba}_{\R_i} )$  is assumed to be concave in $\ol{\ba}_{\R_i}$ and the set $\A_i$ is convex, the Karush Kuhn Tucker (KKT) conditions \cite[Chapter 11]{Boyd} are necessary and sufficient for $\h{\ba}_{\R_i}^*$ to be a maximizer in (\ref{chap5_Cl4_aRi*_argmax}). Thus, for each $i \in \N \tx \la_i \in \Rl_+$ such that, $\h{\ba}_{\R_i}^*$ and $\la_i$ satisfy the KKT conditions given below:
\begin{equation}
 \begin{split}
    \fa j \in \R_i \bs \{i\}, \quad l_{ij}^* - \D_{\ol{a}_j}
        u_i(\ol{\ba}_{\R_i})\mid_{ \ol{\ba}_{\R_i} = \h{\ba}_{\R_i}^* } & = 0, \\
        l_{ii}^*
        - \D_{\ol{a}_i} u_i(\ol{\ba}_{\R_i})\mid_{ \ol{\ba}_{\R_i} = \h{\ba}_{\R_i}^* }
        + \la_i \D_{\ol{a}_i} f_{\A_i}(\ol{a}_i)\mid_{ \ol{a}_i = \h{a}_i^* } & = 0, \\
        \la_i f_{\A_i}(\h{a}_i^*) & = 0.
 \label{chap5_Cl4_KKT_indiv}
 \end{split}
\end{equation}
For each $i \in \N$, adding the KKT condition equations in (\ref{chap5_Cl4_KKT_indiv}) over $k \in \C_i$ results in
\begin{equation}
 \begin{split}
        \sum_{k \in \C_i} l_{ki}^* - \D_{\ol{a}_i} \sum_{k \in \C_i}
        u_k(\ol{\ba}_{\R_k})\mid_{ \ol{\ba}_{\R_k} = \h{\ba}_{\R_k}^* }
        + \la_i \D_{\ol{a}_i} f_{\A_i}(\ol{a}_i)\mid_{ \ol{a}_i = \h{a}_i^* }  
    \ = \ 0. 
 \label{chap5_Cl4_KKT_indiv_sum}
 \end{split}
\end{equation}
From (\ref{chap5_Eq_l_ij}) we have,
\begin{equation}
 \begin{split}
    \sum_{k \in \C_i} l_{ki}^*
    & = \sum_{k \in \C_i} \left( {}^{ \C_{i(\I_{ki}+1)} }\pi_i^*
        - {}^{ \C_{i(\I_{ki}+2)} }\pi_i^* \right) = 0.
 \label{chap5_Cl4_sum_lik_0}
 \end{split}
\end{equation}
Substituting (\ref{chap5_Cl4_sum_lik_0}) in (\ref{chap5_Cl4_KKT_indiv_sum}) we obtain~\fn{The second equality in (\ref{chap5_Cl4_KKT_cent}) is one of the KKT conditions from (\ref{chap5_Cl4_KKT_indiv}).} $\fa i \in \N$,
\begin{equation}
 \begin{split}
    - \D_{\ol{a}_i} \sum_{k \in \C_i}
        u_k(\ol{\ba}_{\R_k})\mid_{ \ol{\ba}_{\R_k} = \h{\ba}_{\R_k}^* }
        + \la_i \D_{\ol{a}_i} f_{\A_i}(\ol{a}_i)\mid_{ \ol{a}_i = \h{a}_i^* } & = 0, \\
        \la_i f_{\A_i}(\h{a}_i^*) & = 0.
 \label{chap5_Cl4_KKT_cent}
 \end{split}
\end{equation}
The conditions in (\ref{chap5_Cl4_KKT_cent}) along with the non-negativity of $\la_i, i \in \N$, specify the KKT conditions (for variable $\h{\ba}_{\N}$) for Problem~($P_C$). Since ($P_C$) is a concave optimization problem, KKT conditions are necessary and sufficient for optimality. As shown in (\ref{chap5_Cl4_KKT_cent}), the action profile $\h{\ba}_{\N}^*$ satisfies these optimality conditions. Furthermore, the tax profile $\h{\bt}_{\N}^*$ satisfies, by its definition, $\sum_{i \in \N} \h{t}_i^* = 0$. Therefore, the NE allocation $(\h{\ba}_{\N}^*, \h{\bt}_{\N}^*)$ is an optimal solution of Problem~($P_C$).
This completes the proof of Claim~\ref{chap5_Cl_NE_optl} and hence, the proof of Theorem~\ref{chap5_Thm_NE_opt}.
\hfill \fbox{}

Claims~\ref{chap5_Cl_fes_sol}--\ref{chap5_Cl_NE_optl} (Theorem~\ref{chap5_Thm_NE_opt}) establish the properties of NE allocations based on the assumption that there exists a NE of the game induced by the game form of Section~\ref{chap5_Subsec_game_form} and users' utility functions (\ref{chap5_Eq_u_i^A}). However, these claims do not guarantee the existence of a NE. This is guaranteed by Theorem~\ref{chap5_Thm_NE_exists} which is proved next in Claims~\ref{chap5_Cl_Cent_Ldl_price} and \ref{chap5_Cl_Ldl_price_NE}.

%%%%%%%%%%%%%%%%%%%%%%%%%%%%%%%%%%%%%%%%%%%%%%%%%%%%%%%%%%%%
\section{Proof of Theorem~2} \label{chap5_Proof_Thm_NE_exists}
%%%%%%%%%%%%%%%%%%%%%%%%%%%%%%%%%%%%%%%%%%%%%%%%%%%%%%%%%%%%

We prove Theorem~\ref{chap5_Thm_NE_exists}  in two steps. In the first step we show that if the centralized problem $(P_C)$ has an optimal action profile $\h{\ba}_{\N}^*$, there exist a set of personalized prices, one for each user $i\in\N$, such that when each $i \in \N$ individually maximizes its own utility taking these prices as given, it obtains $\h{\ba}_{\R_i}^*$ as an optimal action profile. In the second step we show that the optimal action profile $\h{\ba}_{\N}^*$ and the corresponding personalized prices can be used to construct message profiles that are NE of the game induced by the game form of Section~\ref{chap5_Subsec_game_form} and users' utility functions in (\ref{chap5_Eq_u_i^A}).

%%%%%%%%%%%%%%%%%%%%%%%%%%%%%%%%%%%%%%%%%%%%%%%%%%%%%%%%%%%%
\begin{Cl} \label{chap5_Cl_Cent_Ldl_price}
    If Problem~$(P_C)$ has an optimal action profile $\h{\ba}_{\N}^*$, there exist a set of personalized prices $l_{ij}^*, j\in\R_i, i\in\N,$
    such that
    \begin{equation}
        \h{\ba}_{\R_i}^* \in
        \argmax_{ \stackrel{ \h{a}_i \in \A_i }{ \h{a}_j \in \Rl, \, j \in \R_i \bs \{i\} } }
        - \sum_{j \in \R_i} l_{ij}^*\; \h{a}_j + u_i(\h{\ba}_{\R_i}),
        \quad \fa i \in \N.
        \label{chap5_Cl5_a_Ri_argmax}
    \end{equation}
\end{Cl}
%%%%%%%%%%%%%%%%%%%%%%%%%%%%%%%%%%%%%%%%%%%%%%%%%%%%%%%%%%%%
{\bf Proof:}
Suppose $\h{\ba}_{\N}^*$ is an optimal action profile corresponding to Problem~$(P_C)$. Writing the optimization problem $(P_C)$ only in terms of variable $\h{\ba}_{\N}$ gives
\begin{equation}
 \begin{split}
    \h{\ba}_{\N}^* \in
    \argmax_{\h{\ba}_{\N}} & \ \sum_{i \in \N} u_i(\h{\ba}_{\R_i}) \\
    \mb{s.t.}              & \ \ \h{a}_i \in \A_i, \fa i \in \N.
    \label{chap5_Cl5_aN*_argmax}
 \end{split}
\end{equation}
As stated earlier, an optimal solution of Problem~($P_C$) is of the form $(\h{\ba}_{\N}^*, \h{\bt}_{\N})$, where $\h{\ba}_{\N}^*$ is a solution of (\ref{chap5_Cl5_aN*_argmax}) and $\h{\bt}_{\N} \in \Rl^N$ is any tax profile that satisfies (\ref{chap5_Eq_sum_ti_0}). Because KKT conditions are necessary for optimality, the optimal solution in (\ref{chap5_Cl5_aN*_argmax}) must satisfy the KKT conditions. This implies that there exist $\la_i \in \Rl_+, i \in \N,$ such that for each $i \in \N$, $\la_i$ and $\h{\ba}_{\N}^*$ satisfy
\begin{equation}
 \begin{split}
    - \D_{\h{a}_i} \sum_{k \in \C_i}
        u_k(\h{\ba}_{\R_k})\mid_{ \h{\ba}_{\R_k} = \h{\ba}_{\R_k}^* }
        + \la_i \D_{\h{a}_i} f_{\A_i}(\h{a}_i)\mid_{ \h{a}_i = \h{a}_i^* } & = 0, \\
        \la_i f_{\A_i}(\h{a}_i^*) & = 0,
 \label{chap5_Cl5_KKT_cent}
 \end{split}
\end{equation}
where $f_{\A_i}( \cdot )$ is the convex function defined in Claim~\ref{chap5_Cl_NE_optl}.
Defining for each $i \in \N$,
\begin{equation}
 \begin{split}
    l_{ij}^* & :=
    \D_{\h{a}_j} u_i(\h{\ba}_{\R_i})\mid_{ \h{\ba}_{\R_i} = \h{\ba}_{\R_i}^* },
    \qquad j \in \R_i \bs \{i\}, \\
    l_{ii}^* & :=
    \D_{\h{a}_i} u_i(\h{\ba}_{\R_i})\mid_{ \h{\ba}_{\R_i} = \h{\ba}_{\R_i}^* }
    - \la_i \D_{\h{a}_i} f_{\A_i}(\h{a}_i)\mid_{ \h{a}_i = \h{a}_i^* },
  \label{chap5_Cl5_lij_define}
 \end{split}
\end{equation}
we get $\fa i \in \N$,
\begin{equation}
 \begin{split}
     \sum_{k \in \C_i} l_{ki}^* 
   = \D_{\h{a}_i} \sum_{k \in \C_i}
      u_k(\h{\ba}_{\R_k})\mid_{ \h{\ba}_{\R_k} = \h{\ba}_{\R_k}^* }
      - \la_i \D_{\h{a}_i} f_{\A_i}(\h{a}_i)\mid_{ \h{a}_i = \h{a}_i^* }  
 \ =\ 0.
    \label{chap5_Cl5_sum_lki}
 \end{split}
\end{equation}
The second equality in (\ref{chap5_Cl5_sum_lki}) follows from (\ref{chap5_Cl5_KKT_cent}). Furthermore, (\ref{chap5_Cl5_lij_define}) implies that $\fa i \in \N$,
\begin{equation}
 \begin{split}
      \fa j \in \R_i \bs \{i\}, \quad
      l_{ij}^* - \D_{\h{a}_j} u_i(\h{\ba}_{\R_i})\mid_{ \h{\ba}_{\R_i} = \h{\ba}_{\R_i}^* }
    & = 0, \\
      l_{ii}^* - \D_{\h{a}_i} u_i(\h{\ba}_{\R_i})\mid_{ \h{\ba}_{\R_i} = \h{\ba}_{\R_i}^* }
      + \la_i \D_{\h{a}_i} f_{\A_i}(\h{a}_i)\mid_{ \h{a}_i = \h{a}_i^* }
    & = 0.
  \label{chap5_Cl5_KKT_indiv}
 \end{split}
\end{equation}
The equations in (\ref{chap5_Cl5_KKT_indiv}) along with the second equality in (\ref{chap5_Cl5_KKT_cent}) imply that for each $i \in \N$, $\h{\ba}_{\R_i}^*$ and $\la_i$ satisfy the KKT conditions for the following maximization problem:
\begin{equation}
    \max_{ \stackrel{ \h{a}_i \in \A_i }{ \h{a}_j \in \Rl, \, j \in \R_i \bs \{i\} } }
        - \sum_{j \in \R_i} l_{ij}^*\; \h{a}_j + u_i(\h{\ba}_{\R_i})
        \label{chap5_Cl5_max_indiv}
\end{equation}
Because the objective function in (\ref{chap5_Cl5_max_indiv}) is concave (Assumption~\ref{chap5_Ass_util_conc_pvt}), KKT conditions are necessary and sufficient for optimality. Therefore, we conclude from (\ref{chap5_Cl5_KKT_indiv}) and (\ref{chap5_Cl5_KKT_cent}) that,
\begin{equation*}
     \h{\ba}_{\R_i}^* \in
     \argmax_{ \stackrel{ \h{a}_i \in \A_i }{ \h{a}_j \in \Rl, \, j \in \R_i \bs \{i\} } }
     - \sum_{j \in \R_i} l_{ij}^*\; \h{a}_j + u_i(\h{\ba}_{\R_i}),
     \qquad \fa i \in \N.
\end{equation*}
\hfill \fbox{}

%%%%%%%%%%%%%%%%%%%%%%%%%%%%%%%%%%%%%%%%%%%%%%%%%%%%%%%%%%%%
\begin{Cl} \label{chap5_Cl_Ldl_price_NE}
    Let $\h{\ba}_{\N}^*$ be an optimal action profile for Problem~$(P_C)$, let $l_{ij}^*, j\in\R_i, i\in\N,$ be the personalized prices corresponding to $\h{\ba}_{\N}^*$ as defined in Claim~\ref{chap5_Cl_Cent_Ldl_price}, and let $\h{t}_i^* := \sum_{j \in \R_i} l_{ij}^* \h{a}_j^*, i\in\N$. Let $\Bm_i^* := ( \tensor*[^{i}]{\ba}{_{\R_i}^*},  \tensor*[^{i}]{\bpi}{_{\R_i}^*} ), i \in \N,$ be a solution to the following set of relations:
    \begin{eqnarray}
            \frac{1}{|\C_i|}  \sum_{k \in \C_i}   \tensor*[^{k}]{a}{_i^*}
        & =& \h{a}_i^*, \quad i \in \N, \label{chap5_Cl6_NEchar_a_hat} \\
            {}^{ \C_{j(\I_{ij}+1)} }\pi_j^*  -  {}^{ \C_{j(\I_{ij}+2)} }\pi_j^*
        & =& l_{ij}^*,  \quad j \in \R_i, \ i \in \N, \label{chap5_Cl6_NEchar_lij} \\
            \tensor*[^{i}]{\pi}{_j^*}
            \left( \tensor*[^{i}]{a}{_j^*} - {}^{\C_{j(\I_{ij}+1)}}a_j^* \right)^2
        & =& 0, \quad j \in \R_i, \ \ i \in \N, \label{chap5_Cl6_NEchar_pi_a} \\
            \tensor*[^{i}]{\pi}{_j^*}
        & \geq& 0, \quad j \in \R_i, \ \ i \in \N. \label{chap5_Cl6_NEchar_pi_pos}
     \end{eqnarray}
     Then, $\Bm_{\N}^* := (\Bm_1^*, \Bm_2^*, \dots, \Bm_N^*)$ is a NE of the game induced by the game form of Section~\ref{chap5_Subsec_game_form} and the users' utility functions (\ref{chap5_Eq_u_i^A}). Furthermore, for each $i \in \N$, $ \h{a}_i(\Bm_{\C_i}^*) = \h{a}_i^*$, $ l_{ij}(\Bm_{\C_j}^*) = l_{ij}^*,  j \in \R_i$, and $\h{t}_i( (\Bm_{\C_j}^*)_{j \in \R_i} ) = \h{t}_i^*$.
\end{Cl}
%%%%%%%%%%%%%%%%%%%%%%%%%%%%%%%%%%%%%%%%%%%%%%%%%%%%%%%%%%%%
{\bf Proof:}
Note that, the conditions in (\ref{chap5_Cl6_NEchar_a_hat})--(\ref{chap5_Cl6_NEchar_pi_pos}) are necessary for any NE $\Bm_{\N}^*$ of the game induced by the game form of Section~\ref{chap5_Subsec_game_form} and users' utilities (\ref{chap5_Eq_u_i^A}), to result in the allocation $(\h{\ba}_{\N}^*, \h{\bt}_{\N}^*)$ (see (\ref{chap5_Eq_a_hat_i}), (\ref{chap5_Eq_l_ij}) and (\ref{chap5_Cl2_pi_a_0_NE})).  Therefore, the set of solutions of (\ref{chap5_Cl6_NEchar_a_hat})--(\ref{chap5_Cl6_NEchar_pi_pos}), if such a set exists, is a superset of the set of all NE corresponding to the above game that result in $(\h{\ba}_{\N}^*, \h{\bt}_{\N}^*)$. Below we show that the solution set of (\ref{chap5_Cl6_NEchar_a_hat})--(\ref{chap5_Cl6_NEchar_pi_pos}) is in fact exactly the set of all NE that result in $(\h{\ba}_{\N}^*, \h{\bt}_{\N}^*)$.

To prove this, we first show that the set of relations in (\ref{chap5_Cl6_NEchar_a_hat})--(\ref{chap5_Cl6_NEchar_pi_pos}) do have a solution. Notice that (\ref{chap5_Cl6_NEchar_a_hat}) and (\ref{chap5_Cl6_NEchar_pi_a}) are satisfied by setting for each $i \in \N$, $\tensor*[^{k}]{a}{_i^*} = \h{a}_i^* \fa k \in \C_i$. Notice also that for each $j \in \N$, the sum over $i \in \C_j$ of the right hand side of (\ref{chap5_Cl6_NEchar_lij}) is 0. Therefore, for each $j \in \N$, (\ref{chap5_Cl6_NEchar_lij}) has a solution in ${}^{i}\pi_j^*, i \in \C_j$. Furthermore, for any solution ${}^{i}\pi_j^*, i \in \C_j, j \in \N$, of (\ref{chap5_Cl6_NEchar_lij}), ${}^{i}\pi_j^* + c, i \in \C_j, j \in \N$, where $c$ is some constant, is also a solution of (\ref{chap5_Cl6_NEchar_lij}). Consequently, by appropriately choosing $c$, we can select a solution of (\ref{chap5_Cl6_NEchar_lij}) such that (\ref{chap5_Cl6_NEchar_pi_pos}) is satisfied.

It is clear from the above discussion that (\ref{chap5_Cl6_NEchar_a_hat})--(\ref{chap5_Cl6_NEchar_pi_pos}) have multiple solutions. We now show that the set of solutions $\Bm_{\N}^*$ of (\ref{chap5_Cl6_NEchar_a_hat})--(\ref{chap5_Cl6_NEchar_pi_pos}) is the set of NE that result in $(\h{\ba}_{\N}^*, \h{\bt}_{\N}^*)$. From Claim~\ref{chap5_Cl_Cent_Ldl_price}, (\ref{chap5_Cl5_a_Ri_argmax}) can be equivalently written as
\begin{equation}
 \begin{split}
         \h{\ba}_{\R_i}^*
     %% & = \argmax_{ \h{\ba}_{\R_i} \in \Rl^{|\R_i|} }
     %%    - \sum_{j \in \R_i} l_{ij}^*\; \h{a}_j + u_i(\h{\ba}_{\R_i})
     %%    - \left[ \frac{1 - I_{\A_i}(a_i)}{I_{\A_i}(a_i)} \right] \\
     & \in \argmax_{ \h{\ba}_{\R_i} \in \Rl^{|\R_i|} }
         u_i^A\Big( \h{\ba}_{\R_i}, \sum_{j \in \R_i} l_{ij}^*\; \h{a}_j \Big),
        \qquad i \in \N.
        \label{chap5_Cl6_a_Ri_argmax}
 \end{split}
\end{equation}
Substituting $\h{a}_j |\C_j| - \sum_{k \in \C_j \bs \{i\} } {}^{k}a_j^* = {}^{i}a_j$ for each $j \in \R_i$, $i \in \N$, in (\ref{chap5_Cl6_a_Ri_argmax}) we obtain
\begin{equation}
 \begin{split}
        \tensor*[^{i}]{\ba}{_{\R_i}^*}  \in 
        \argmax_{ \tensor*[^{i}]{\ba}{_{\R_i}} \in \Rl^{|\R_i|} } u_i^A\bigg(
       \Big( \frac{1}{|\C_j|} \big( {}^{i}a_j 
        + \sum_{k \in \C_j \bs \{i\} } {}^{k}a_j^* \big) \Big)_{j \in \R_i}, 
       \sum_{j \in \R_i} l_{ij}^*\; \frac{1}{|\C_j|} \big( {}^{i}a_j
        + \sum_{k \in \C_j \bs \{i\} } {}^{k}a_j^* \big) \bigg),  
       \ i \in \N. 
        \label{chap5_Cl6_ia_Ri_argmax}
 \end{split}
\end{equation}
Because of (\ref{chap5_Cl6_NEchar_pi_a}), (\ref{chap5_Cl6_ia_Ri_argmax}) also implies that
\begin{equation}
 \begin{split}
       & ( \tensor*[^{i}]{\ba}{_{\R_i}^*},  \tensor*[^{i}]{\bpi}{_{\R_i}^*} ) \in 
         \argmax_{ ( \tensor*[^{i}]{\ba}{_{\R_i}}, \tensor*[^{i}]{\bpi}{_{\R_i}} )
         \in \Rl^{|\R_i|} \times \Rl_+^{|\R_i|} }
         u_i^A\bigg( \Big( \h{a}_j \big(
         ( {}^{i}\ba_{\R_i}, {}^{i}\bpi_{\R_i} ), \; \Bm_{\C_j}^* / i \big)
         \Big)_{j \in \R_i}, \\
       & \sum_{j \in \R_i} l_{ij}^*\;
         \h{a}_j \big( ( {}^{i}\ba_{\R_i}, {}^{i}\bpi_{\R_i} ), \; \Bm_{\C_j}^* / i \big)
         -  \sum_{j \in \R_i}  {}^{ \C_{j(\I_{ij}+1)} }\pi_j^*
         \left( {}^{\C_{j(\I_{ij}+1)}}a_j^*
         - {}^{\C_{j(\I_{ij}+2)}}a_j^* \right)^2
         \bigg), \  i \in \N.
        \label{chap5_Cl6_iaRi_ipiRi_argmax}
 \end{split}
\end{equation}
Furthermore, since $u_i^A$ is strictly decreasing in the tax (see (\ref{chap5_Eq_u_i^A})), (\ref{chap5_Cl6_iaRi_ipiRi_argmax}) also implies the following:
\begin{equation}
 \begin{split}
         ( \tensor*[^{i}]{\ba}{_{\R_i}^*}, \tensor*[^{i}]{\bpi}{_{\R_i}^*} ) \in 
       & \hs{-0.3in}\argmax_{( \tensor*[^{i}]{\ba}{_{\R_i}}, \tensor*[^{i}]{\bpi}{_{\R_i}} )
         \in \Rl^{|\R_i|} \times \Rl_+^{|\R_i|} }
         u_i^A\bigg( \Big( \h{a}_j \big(
         ( {}^{i}\ba_{\R_i}, {}^{i}\bpi_{\R_i} ),  \Bm_{\C_j}^* / i \big)
         \Big)_{j \in \R_i}, 
         \sum_{j \in \R_i} l_{ij}^*
         \h{a}_j \big( ( {}^{i}\ba_{\R_i}, {}^{i}\bpi_{\R_i} ), \Bm_{\C_j}^* / i \big) \\
        & \hs{-0.3in} +  \sum_{j \in \R_i}  {}^{i}\pi_j
         \left( {}^{i}a_j - {}^{\C_{j(\I_{ij}+1)}}a_j^* \right)^2 
         -  \sum_{j \in \R_i}  {}^{ \C_{j(\I_{ij}+1)} }\pi_j^*
         \left( {}^{\C_{j(\I_{ij}+1)}}a_j^*
         - {}^{\C_{j(\I_{ij}+2)}}a_j^* \right)^2
         \bigg),
         \ \quad i \in \N.
        \label{chap5_Cl6_iaRi_ipiRi_NE_argmax}
 \end{split}
\end{equation}
Eq.~(\ref{chap5_Cl6_iaRi_ipiRi_NE_argmax}) implies that, if the message exchange and allocation is done according to the game form presented in Section~\ref{chap5_Subsec_game_form}, then user $i, i \in \N,$ maximizes its utility at $\Bm_i^*$ when all other users $j \in \N \bs \{i\}$ choose their respective messages $\Bm_j^*, j \in \N \bs \{i\}$. This, in turn, implies that a message profile $\Bm_{\N}^*$ that is a solution to (\ref{chap5_Cl6_NEchar_a_hat})--(\ref{chap5_Cl6_NEchar_pi_pos}) is a NE of the game induced by the aforementioned game form and the users' utilities (\ref{chap5_Eq_u_i^A}). Furthermore, it follows from (\ref{chap5_Cl6_NEchar_a_hat})--(\ref{chap5_Cl6_NEchar_pi_pos}) that the allocation at $\Bm_{\N}^*$ is
\begin{equation}
 \begin{split}
         \h{a}_i(\Bm_{\C_i}^*)
  =  & \frac{1}{|\C_i|}  \sum_{k \in \C_i} \tensor*[^{k}]{a}{_i^*}
         \ = \ \h{a}_i^*, \hs{1.65in} i \in \N, \\
         l_{ij}(\Bm_{\C_j}^*)
  =  & {}^{ \C_{j(\I_{ij}+1)} }\pi_j^*  -  {}^{ \C_{j(\I_{ij}+2)} }\pi_j^* 
         \ = \ l_{ij}^*,  \qquad j \in \R_i, \ i \in \N, \\
 \hs{-0.3in}        \h{t}_i\big( (\Bm_{\C_j}^*)_{j \in \R_i} \big)
 \! = \! & \sum_{j \in \R_i} l_{ij} (\Bm_{\C_j}^*) \h{a}_j(\Bm_{\C_j}^*)
         \!+\! \tensor*[^{i}]{\pi}{_j^*} \left( \tensor*[^{i}]{a}{_j^*} - {}^{\C_{j(\I_{ij}+1)}}a_j^* \right)^2 
        \!-\!  {}^{\C_{j(\I_{ij}+1)}}\pi_j^* 
         \left( {}^{\C_{j(\I_{ij}+1)}}a_j^* - {}^{\C_{j(\I_{ij}+2)}}a_j^* \right)^2 \\
 \!  = \! & \sum_{j \in \R_i} l_{ij}^* \h{a}_i^* \ = \  \h{t}_i^*,  \hs{1.8in} i \in \N.
 \label{chap5_Cl6_NE_cent}
 \end{split}
\end{equation}
From (\ref{chap5_Cl6_NE_cent}) it follows that the set of solutions $\Bm_{\N}^*$ of (\ref{chap5_Cl6_NEchar_a_hat})--(\ref{chap5_Cl6_NEchar_pi_pos}) is exactly the set of NE that result in $(\h{\ba}_{\N}^*, \h{\bt}_{\N}^*)$.
This completes the proof of Claim~\ref{chap5_Cl_Ldl_price_NE} and hence the proof of Theorem~\ref{chap5_Thm_NE_exists}.
\hfill \fbox{}

%%%%%%%%%%%%%%%%%%%%%%%%%%%%%%%%%%%%%%%%%%%%%%%%%%%%%%%%%%%%
\section{An example to illustrate properties of the game form} \label{Apx_illus_ex}
%%%%%%%%%%%%%%%%%%%%%%%%%%%%%%%%%%%%%%%%%%%%%%%%%%%%%%%%%%%%

Consider a system with three users where each user's action affects the utilities of all three users. Let the sets of feasible actions of the users be $\A_1 = \A_2 = \A_3 = [0,1]$. Suppose that the users' utilities are:
  \begin{equation}
  	\begin{split}
  		u_1(\h{a}_1, \h{a}_2, \h{a}_3) & = \ \ \h{a}_1^{\al_1} - \h{a}_2^{\Be_2} - \h{a}_3^{\Be_3}, 
  			\quad \h{a}_1 \in \A_1, \ (\h{a}_2, \h{a}_3) \in \Rl^2, \\
  		u_2(\h{a}_1, \h{a}_2, \h{a}_3) & = - \h{a}_1^{\Be_1} + \h{a}_2^{\al_2} - \h{a}_3^{\Be_3}, 
  			\quad \h{a}_2 \in \A_2, \ (\h{a}_1, \h{a}_3) \in \Rl^2, \\
  		u_3(\h{a}_1, \h{a}_2, \h{a}_3) & = - \h{a}_1^{\Be_1} - \h{a}_2^{\Be_2} + \h{a}_3^{\al_3}, 
  			\quad \h{a}_3 \in \A_3, \ (\h{a}_1, \h{a}_2) \in \Rl^2, \\
  		 \mb{where}, & \quad \ \al_i \in (0,1), \ \Be_i \in (1,\infty) \ \fa i\in\{1,2,3\}.
  		\label{u_i_def}
  	\end{split}
  \end{equation}
 Because of the parameters $\al_i, \Be_i, i\in\{1,2,3\}$, the utility functions in (\ref{u_i_def}) are strictly concave over their domains. Therefore, there is a unique optimum action profile $\h{\ba}^* = (\h{a}_1^*, \h{a}_2^*,\h{a}_3^*)$ corresponding to the centralized problem $(P_C)$ defined in (\ref{chap5_Eq_Pc_obj2}). This optimum action profile $\h{\ba}^*$ is the solution to $\D_{\h{\ba}} (\sum_{i\in\{1,2,3\}} u_i(\h{a}_1, \h{a}_2, \h{a}_3)) = 0$.  
Computing the gradient in this equation we obtain,
 \begin{eqnarray}
 		\left[
 		\begin{array}{l}
 			\al_1 \,\h{a}_1^{*^{(\al_1-1)}} - 2 \,\Be_1 \,\h{a}_1^{*^{(\Be_1-1)}} \\
 			\al_2 \,\h{a}_2^{*^{(\al_2-1)}} - 2 \,\Be_2 \,\h{a}_2^{*^{(\Be_2-1)}} \\
 			\al_3 \,\h{a}_3^{*^{(\al_3-1)}} - 2 \,\Be_3 \,\h{a}_3^{*^{(\Be_3-1)}}
 		\end{array}
 		\right]
 		& = & 
 		0 \nnb \\
 		\Raro 
 		\left[
 		\begin{array}{l}
 			{\h{a}_1^*} \\
 			{\h{a}_2^*} \\
 			{\h{a}_3^*}
 		\end{array}
 		\right]
 		& = &
  		\left[
 		\begin{array}{l}
 			\big(\frac{\al_1}{2 \Be_1}\big)^{1/(\Be_1 - \al_1)} \\
 			\big(\frac{\al_2}{2 \Be_2}\big)^{1/(\Be_2 - \al_2)} \\
 			\big(\frac{\al_3}{2 \Be_3}\big)^{1/(\Be_3 - \al_3)}
 		\end{array}
 		\right]	
 		\label{a*_cent}	
 \end{eqnarray}
     Note that because $\al_i < \Be_i \fa i\in\{1,2,3\},$ the action profile $\h{\ba}^*$ in (\ref{a*_cent}) lies in $\A_1\times\A_2\times\A_3$ and hence is feasible. Equation (\ref{a*_cent}) thus gives the optimum centralized action profile. 
     \\
     
    We now show the properties of the game form presented in Section~\ref{chap5_Subsec_game_form}.
    
    We first show that the game induced by this game form under the above problem instance has a Nash equilibrium.       Consider the following message profile of the users:
     \begin{equation}
     	\begin{split}
     		\Bm_1^* = 
     		\big( (\tensor*[^{1}]{a}{_{1}^*}, \tensor*[^{1}]{a}{_{2}^*}, \tensor*[^{1}]{a}{_{3}^*}), \ 
     					(\tensor*[^{1}]{\pi}{_{1}^*}, \tensor*[^{1}]{\pi}{_{2}^*}, \tensor*[^{1}]{\pi}{_{3}^*})
     		\big)
     		& = 
     		\big( (\h{a}_1^*, \h{a}_2^*, \h{a}_3^*),\ 
     					(2 \Be_1 \h{a}_1^{*^{(\Be_1-1)}}, \Be_2 \h{a}_2^{*^{(\Be_2-1)}}, 
     					 3 \Be_3 \h{a}_3^{*^{(\Be_3-1)}})     		
     		\big), \\
     		\Bm_2^* = 
     		\big( (\tensor*[^{2}]{a}{_{1}^*}, \tensor*[^{2}]{a}{_{2}^*}, \tensor*[^{2}]{a}{_{3}^*}), \ 
     					(\tensor*[^{2}]{\pi}{_{1}^*}, \tensor*[^{2}]{\pi}{_{2}^*}, \tensor*[^{2}]{\pi}{_{3}^*})
     		\big)
     		& = 
     		\big( (\h{a}_1^*, \h{a}_2^*, \h{a}_3^*),\ 
     					(3 \Be_1 \h{a}_1^{*^{(\Be_1-1)}}, 2 \Be_2 \h{a}_2^{*^{(\Be_2-1)}}, 
     					   \Be_3 \h{a}_3^{*^{(\Be_3-1)}})     		
     		\big), \\
     		\Bm_3^* = 
     		\big( (\tensor*[^{3}]{a}{_{1}^*}, \tensor*[^{3}]{a}{_{2}^*}, \tensor*[^{3}]{a}{_{3}^*}), \ 
     					(\tensor*[^{3}]{\pi}{_{1}^*}, \tensor*[^{3}]{\pi}{_{2}^*}, \tensor*[^{3}]{\pi}{_{3}^*})
     		\big)
     		& = 
     		\big( (\h{a}_1^*, \h{a}_2^*, \h{a}_3^*),\ 
     					( \Be_1 \h{a}_1^{*^{(\Be_1-1)}}, 3 \Be_2 \h{a}_2^{*^{(\Be_2-1)}}, 
     					 2 \Be_3 \h{a}_3^{*^{(\Be_3-1)}})     		
     		\big). 
     		\label{msg_prof}
     	\end{split}
     \end{equation} 
     The vector $(\h{a}_1^*, \h{a}_2^*, \h{a}_3^*)$ on the right hand side (RHS) of (\ref{msg_prof}) is the optimum centralized action profile computed in (\ref{a*_cent}), and $\Be_i, i\in\{1,2,3\}$, are the parameters of the utility functions in (\ref{u_i_def}). Because $\Be_i \in (1,\infty) \fa i\in\{1,2,3\}$, and $(\h{a}_1^*, \h{a}_2^*, \h{a}_3^*) \in \A_1\times\A_2\times\A_3$, (\ref{msg_prof}) implies that $\tensor*[^{i}]{\pi}{_{j}^*}\in\Rl_+ \fa i,j \in \{1,2,3\}$. Hence the message profile given in (\ref{msg_prof}) lies in the message space (defined by (\ref{chap5_Eq_mi}), (\ref{chap5_Eq_i_p_Ai})) of the proposed game form.
     
%    According to our mechanism, the outcome corresponding to the above message profile is computed as follows:
%    \begin{equation}
%    	\begin{split}
%    		\h{a}_1()
%    	\end{split}
%    \end{equation}

We will now show that the above message profile is a Nash equilibrium of the game induced by the proposed game form and the users' utility functions in (\ref{u_i_def}). To this end we will show that, 
\begin{eqnarray}
%	\begin{split}
		\Bm_1^* &\in& \argmax_{\Bm_1 \in \Rl^3\times\Rl_+^3} 
				\left\{ u_1\big(\big(\h{a}_i(\Bm_1, \Bm_2^*, \Bm_3^*)\big)_{i\in\{1,2,3\}}\big) - \h{t}_1(\Bm_1, \Bm_2^*, \Bm_3^*) \right\}, \label{NE_argmax_cond_1} \\
		\Bm_2^* &\in& \argmax_{\Bm_2 \in \Rl^3\times\Rl_+^3} 
				\left\{ u_2\big(\big(\h{a}_i(\Bm_1^*, \Bm_2, \Bm_3^*)\big)_{i\in\{1,2,3\}}\big) - \h{t}_2(\Bm_1^*, \Bm_2, \Bm_3^*) \right\}, \label{NE_argmax_cond_2}\\
		\Bm_3^* &\in& \argmax_{\Bm_3 \in \Rl^3\times\Rl_+^3} 
				\left\{ u_3\big(\big(\h{a}_i(\Bm_1^*, \Bm_2^*, \Bm_3)\big)_{i\in\{1,2,3\}}\big) - \h{t}_3(\Bm_1^*, \Bm_2^*, \Bm_3) \right\},
				\label{NE_argmax_cond_3}
%	\end{split}	
\end{eqnarray} 
where, $u_i, i\in\{1,2,3\}$, are the utility functions defined in (\ref{u_i_def}), and $\h{a}_i, \h{t}_i, i\in\{1,2,3\}$, are the action and tax functions of the game form defined by equations (\ref{chap5_Eq_a_hat_i})--(\ref{chap5_Eq_l_ij}). %\textcolor{red}{Check exact place}.\\
Below we present a derivation for (\ref{NE_argmax_cond_1}). The conditions (\ref{NE_argmax_cond_2}) and (\ref{NE_argmax_cond_3}) can be derived similarly.\\

Let the indexing of the users for tax computation be the same as their original indexing, i.e. users $1,2,3$ are labeled as $1,2,3$ in all the cycles $\C_1, \C_2, \C_3$ (consisting of all three users) used in tax functions. Then, using equations (\ref{chap5_Eq_t_hat_i}), (\ref{chap5_Eq_l_ij}), the tax of user 1 at message profile $(\Bm_1, \Bm_2^*, \Bm_3^*)$ is: %\textcolor{red}{check exact numbers}  
\begin{equation}
\begin{split}
	\h{t}_1(\Bm_1, \Bm_2^*, \Bm_3^*) = & \ \ \ 
	\big( \tensor*[^{2}]{\pi}{_{1}^*} - \tensor*[^{3}]{\pi}{_{1}^*} \big) \; 
	 \frac{1}{3} \big(\tensor*[^{1}]{a}{_{1}} + \tensor*[^{2}]{a}{_{1}^*} + \tensor*[^{3}]{a}{_{1}^*} \big) \\
	 & + 
	 \big( \tensor*[^{2}]{\pi}{_{2}^*} - \tensor*[^{3}]{\pi}{_{2}^*} \big) \; 
	 \frac{1}{3} \big(\tensor*[^{1}]{a}{_{2}} + \tensor*[^{2}]{a}{_{2}^*} + \tensor*[^{3}]{a}{_{2}^*} \big)  
   + \big( \tensor*[^{2}]{\pi}{_{3}^*} - \tensor*[^{3}]{\pi}{_{3}^*} \big) \; 
	 \frac{1}{3} \big(\tensor*[^{1}]{a}{_{3}} + \tensor*[^{2}]{a}{_{3}^*} + \tensor*[^{3}]{a}{_{3}^*} \big) \\
	 & + 
	 \tensor*[^{1}]{\pi}{_{1}} \big(\tensor*[^{1}]{a}{_{1}} - \tensor*[^{2}]{a}{_{1}^*} \big)^2 +
	 \tensor*[^{1}]{\pi}{_{2}} \big(\tensor*[^{1}]{a}{_{2}} - \tensor*[^{2}]{a}{_{2}^*} \big)^2  + 
	 \tensor*[^{1}]{\pi}{_{3}} \big(\tensor*[^{1}]{a}{_{3}} - \tensor*[^{2}]{a}{_{3}^*} \big)^2 \\
	 & -
	 \tensor*[^{2}]{\pi}{_{1}^*} \big(\tensor*[^{2}]{a}{_{1}^*} - \tensor*[^{3}]{a}{_{1}^*} \big)^2 -
	 \tensor*[^{2}]{\pi}{_{2}^*} \big(\tensor*[^{2}]{a}{_{2}^*} - \tensor*[^{3}]{a}{_{2}^*} \big)^2 -
	 \tensor*[^{2}]{\pi}{_{3}^*} \big(\tensor*[^{2}]{a}{_{3}^*} - \tensor*[^{3}]{a}{_{3}^*} \big)^2.
	 \label{tax1_expanded} 
	\end{split}
\end{equation} 
%\begin{equation}
%\begin{split}
%	& \h{t}_1(\Bm_1, \Bm_2^*, \Bm_3^*) = \\
%	& 
%	 \big(3 \Be_1 \h{a}_1^{*^{(\Be_1-1)}} \!\!\!\!\!\!- \Be_1 \h{a}_1^{*^{(\Be_1-1)}}\big) \; 
%	 \frac{1}{3} \big(\tensor*[^{1}]{a}{_{1}} + \tensor*[^{2}]{a}{_{1}^*} + \tensor*[^{3}]{a}{_{1}^*} \big) + 
%	 \big(2 \Be_2 \h{a}_2^{*^{(\Be_2-1)}} \!\!\!\!\!\!- 3 \Be_2 \h{a}_2^{*^{(\Be_2-1)}}\big) \; 
%	 \frac{1}{3} \big(\tensor*[^{1}]{a}{_{2}} + \tensor*[^{2}]{a}{_{2}^*} + \tensor*[^{3}]{a}{_{2}^*} \big)  \\
%   & 
%   + \big(\Be_3 \h{a}_3^{*^{(\Be_3-1)}} \!\!\!\!\!\! - 2 \Be_3 \h{a}_3^{*^{(\Be_3-1)}}\big) \; 
%	 \frac{1}{3} \big(\tensor*[^{1}]{a}{_{3}} + \tensor*[^{2}]{a}{_{3}^*} + \tensor*[^{3}]{a}{_{3}^*} \big) + 
%	 \tensor*[^{1}]{\pi}{_{1}} \big(\tensor*[^{1}]{a}{_{1}} - \tensor*[^{2}]{a}{_{1}^*} \big)^2 +
%	 \tensor*[^{1}]{\pi}{_{2}} \big(\tensor*[^{1}]{a}{_{2}} - \tensor*[^{2}]{a}{_{2}^*} \big)^2  \\
%	 &
%	 + \tensor*[^{1}]{\pi}{_{3}} \big(\tensor*[^{1}]{a}{_{3}} - \tensor*[^{2}]{a}{_{3}^*} \big)^2 -
%	 \tensor*[^{2}]{\pi}{_{1}} \big(\tensor*[^{2}]{a}{_{1}^*} - \tensor*[^{3}]{a}{_{1}^*} \big)^2 -
%	 \tensor*[^{2}]{\pi}{_{2}} \big(\tensor*[^{2}]{a}{_{2}^*} - \tensor*[^{3}]{a}{_{2}^*} \big)^2 -
%	 \tensor*[^{2}]{\pi}{_{3}} \big(\tensor*[^{2}]{a}{_{3}^*} - \tensor*[^{3}]{a}{_{3}^*} \big)^2.
%	 \label{tax1_expanded} 
%	\end{split}
%\end{equation} 
The last three terms on the RHS of (\ref{tax1_expanded}) are zero from the definition of $\tensor*[^{i}]{a}{_{j}^*}, i\in\{2,3\}, j\in\{1,2,3\}$, in (\ref{msg_prof}). Substituting the remaining terms of $\h{t}_1$ in the argument of (\ref{NE_argmax_cond_1}) and expanding $u_1$ we obtain,
\begin{equation}
	\begin{split}
		& u_1\big(\big(\h{a}_i(\Bm_1, \Bm_2^*, \Bm_3^*)\big)_{i\in\{1,2,3\}}\big) - \h{t}_1(\Bm_1, \Bm_2^*, \Bm_3^*) = \\
		& \Big(\frac{1}{3} \big(\tensor*[^{1}]{a}{_{1}} + \tensor*[^{2}]{a}{_{1}^*} +
		   \tensor*[^{3}]{a}{_{1}^*}\big)\Big)^{\al_1} 
		  -\Big(\frac{1}{3} \big(\tensor*[^{1}]{a}{_{2}} + \tensor*[^{2}]{a}{_{2}^*} +
		   \tensor*[^{3}]{a}{_{2}^*}\big)\Big)^{\Be_2} 
		  -\Big(\frac{1}{3} \big(\tensor*[^{1}]{a}{_{3}} + \tensor*[^{2}]{a}{_{3}^*} +
		   \tensor*[^{3}]{a}{_{3}^*}\big)\Big)^{\Be_3} 
		  - \frac{1}{3} \big( \tensor*[^{2}]{\pi}{_{1}^*} - \tensor*[^{3}]{\pi}{_{1}^*} \big) \\
	& 	 \big(\tensor*[^{1}]{a}{_{1}} + \tensor*[^{2}]{a}{_{1}^*} + \tensor*[^{3}]{a}{_{1}^*} \big) 
			- \frac{1}{3}  \big( \tensor*[^{2}]{\pi}{_{2}^*} - \tensor*[^{3}]{\pi}{_{2}^*} \big)
			 \big(\tensor*[^{1}]{a}{_{2}} + \tensor*[^{2}]{a}{_{2}^*} + \tensor*[^{3}]{a}{_{2}^*} \big)  
     - \frac{1}{3} \big( \tensor*[^{2}]{\pi}{_{3}^*} - \tensor*[^{3}]{\pi}{_{3}^*} \big)
       \big(\tensor*[^{1}]{a}{_{3}} + \tensor*[^{2}]{a}{_{3}^*} + \tensor*[^{3}]{a}{_{3}^*} \big) \\
   & 
      - \tensor*[^{1}]{\pi}{_{1}} \big(\tensor*[^{1}]{a}{_{1}} - \tensor*[^{2}]{a}{_{1}^*} \big)^2 
      - \tensor*[^{1}]{\pi}{_{2}} \big(\tensor*[^{1}]{a}{_{2}} - \tensor*[^{2}]{a}{_{2}^*} \big)^2  
	    - \tensor*[^{1}]{\pi}{_{3}} \big(\tensor*[^{1}]{a}{_{3}} - \tensor*[^{2}]{a}{_{3}^*} \big)^2.
	    \label{u1_t1_expanded}
    \end{split}
\end{equation} 
%\begin{equation}
%	\begin{split}
%		& u_1\big(\big(\h{a}_i(\Bm_1, \Bm_2^*, \Bm_3^*)\big)_{i\in\{1,2,3\}}\big) - \h{t}_1(\Bm_1, \Bm_2^*, \Bm_3^*) = \\
%		& \Big(\frac{1}{3} \big(\tensor*[^{1}]{a}{_{1}} + \tensor*[^{2}]{a}{_{1}^*} +
%		   \tensor*[^{3}]{a}{_{1}^*}\big)\Big)^{\al_1} 
%		  -\Big(\frac{1}{3} \big(\tensor*[^{1}]{a}{_{2}} + \tensor*[^{2}]{a}{_{2}^*} +
%		   \tensor*[^{3}]{a}{_{2}^*}\big)\Big)^{\Be_2} 
%		  -\Big(\frac{1}{3} \big(\tensor*[^{1}]{a}{_{3}} + \tensor*[^{2}]{a}{_{3}^*} +
%		   \tensor*[^{3}]{a}{_{3}^*}\big)\Big)^{\Be_3} \\
%		& - \frac{1}{3} \big(3 \Be_1 \h{a}_1^{*^{(\Be_1-1)}} \!\!\!\!\!\!- \Be_1 \h{a}_1^{*^{(\Be_1-1)}}\big) 
%			 \big(\tensor*[^{1}]{a}{_{1}} + \tensor*[^{2}]{a}{_{1}^*} + \tensor*[^{3}]{a}{_{1}^*} \big) 
%			- \frac{1}{3}  \big(2 \Be_2 \h{a}_2^{*^{(\Be_2-1)}} \!\!\!\!\!\!- 3 \Be_2 \h{a}_2^{*^{(\Be_2-1)}}\big)
%			 \big(\tensor*[^{1}]{a}{_{2}} + \tensor*[^{2}]{a}{_{2}^*} + \tensor*[^{3}]{a}{_{2}^*} \big)  \\
%   &  - \frac{1}{3} \big(\Be_3 \h{a}_3^{*^{(\Be_3-1)}} \!\!\!\!\!\! - 2 \Be_3 \h{a}_3^{*^{(\Be_3-1)}}\big)
%       \big(\tensor*[^{1}]{a}{_{3}} + \tensor*[^{2}]{a}{_{3}^*} + \tensor*[^{3}]{a}{_{3}^*} \big)
%      - \tensor*[^{1}]{\pi}{_{1}} \big(\tensor*[^{1}]{a}{_{1}} - \tensor*[^{2}]{a}{_{1}^*} \big)^2 
%      - \tensor*[^{1}]{\pi}{_{2}} \big(\tensor*[^{1}]{a}{_{2}} - \tensor*[^{2}]{a}{_{2}^*} \big)^2  \\
%	 &
%	    - \tensor*[^{1}]{\pi}{_{3}} \big(\tensor*[^{1}]{a}{_{3}} - \tensor*[^{2}]{a}{_{3}^*} \big)^2.
%	    \label{u1_t1_expanded}
%    \end{split}
%\end{equation} 
Note that the last three terms on the RHS of (\ref{u1_t1_expanded}) are non positive for all messages $\big((\tensor*[^{1}]{a}{_{1}}, \tensor*[^{1}]{a}{_{2}}, \tensor*[^{1}]{a}{_{3}}),$ $(\tensor*[^{1}]{\pi}{_{1}}, \tensor*[^{1}]{\pi}{_{2}}, \tensor*[^{1}]{\pi}{_{3}}) \big) \in \Rl^3\times\Rl_+^3$. Furthermore, the price proposals $(\tensor*[^{1}]{\pi}{_{1}}, \tensor*[^{1}]{\pi}{_{2}}, \tensor*[^{1}]{\pi}{_{3}})$ of user 1 affect only these three terms on the RHS of (\ref{u1_t1_expanded}). Therefore, in
any message $\Bm_1$ that maximizes (\ref{u1_t1_expanded}), if $\tensor*[^{1}]{a}{_{1}} \neq \tensor*[^{2}]{a}{_{1}^*}$  then $\tensor*[^{1}]{\pi}{_{1}} = 0$  because $\tensor*[^{1}]{\pi}{_{1}}$ can independently control the value of the quadratic term $\tensor*[^{1}]{\pi}{_{1}} \big(\tensor*[^{1}]{a}{_{1}} - \tensor*[^{2}]{a}{_{1}^*} \big)^2$.  
Similarly, if $\tensor*[^{1}]{a}{_{2}} \neq \tensor*[^{2}]{a}{_{2}^*}$ and $\tensor*[^{1}]{a}{_{3}} \neq \tensor*[^{2}]{a}{_{3}^*}$, then $\tensor*[^{1}]{\pi}{_{2}} = 0$ and $\tensor*[^{1}]{\pi}{_{3}} = 0$. Consequently, the last three terms on the RHS of (\ref{u1_t1_expanded}) must be zero at any message that maximizes this expression. 
As a result, the maximization of (\ref{u1_t1_expanded}) with respect to (w.r.t) $\Bm_1$ is equivalent to the maximization of the following function $g( \tensor*[^{1}]{a}{_{1}}, \tensor*[^{1}]{a}{_{2}}, \tensor*[^{1}]{a}{_{3}})$ w.r.t. $(\tensor*[^{1}]{a}{_{1}}, \tensor*[^{1}]{a}{_{2}}, \tensor*[^{1}]{a}{_{3}})$:
\begin{equation}
	\begin{split}
		& g( \tensor*[^{1}]{a}{_{1}}, \tensor*[^{1}]{a}{_{2}}, \tensor*[^{1}]{a}{_{3}})  := \\
		& \Big(\frac{1}{3} \big(\tensor*[^{1}]{a}{_{1}} + \tensor*[^{2}]{a}{_{1}^*} +
		   \tensor*[^{3}]{a}{_{1}^*}\big)\Big)^{\al_1} 
		  -\Big(\frac{1}{3} \big(\tensor*[^{1}]{a}{_{2}} + \tensor*[^{2}]{a}{_{2}^*} +
		   \tensor*[^{3}]{a}{_{2}^*}\big)\Big)^{\Be_2} 
		  -\Big(\frac{1}{3} \big(\tensor*[^{1}]{a}{_{3}} + \tensor*[^{2}]{a}{_{3}^*} +
		   \tensor*[^{3}]{a}{_{3}^*}\big)\Big)^{\Be_3} 
		  - \frac{1}{3} \big( \tensor*[^{2}]{\pi}{_{1}^*} - \tensor*[^{3}]{\pi}{_{1}^*} \big) \\
	& 	 \big(\tensor*[^{1}]{a}{_{1}} + \tensor*[^{2}]{a}{_{1}^*} + \tensor*[^{3}]{a}{_{1}^*} \big) 
			- \frac{1}{3}  \big( \tensor*[^{2}]{\pi}{_{2}^*} - \tensor*[^{3}]{\pi}{_{2}^*} \big)
			 \big(\tensor*[^{1}]{a}{_{2}} + \tensor*[^{2}]{a}{_{2}^*} + \tensor*[^{3}]{a}{_{2}^*} \big)  
     - \frac{1}{3} \big( \tensor*[^{2}]{\pi}{_{3}^*} - \tensor*[^{3}]{\pi}{_{3}^*} \big)
       \big(\tensor*[^{1}]{a}{_{3}} + \tensor*[^{2}]{a}{_{3}^*} + \tensor*[^{3}]{a}{_{3}^*} \big)
       \label{g_a}
	\end{split}
\end{equation}
Because of the values of the parameters $\al_1$, $\Be_2$ and $\Be_3$, the function $g( \tensor*[^{1}]{a}{_{1}}, \tensor*[^{1}]{a}{_{2}}, \tensor*[^{1}]{a}{_{3}})$ is strictly concave. 
%in $(\tensor*[^{1}]{a}{_{1}}, \tensor*[^{1}]{a}{_{2}}, \tensor*[^{1}]{a}{_{3}})$. 
Therefore, the maximizer of the function is the solution to  $\D_{(\tensor*[^{1}]{a}{_{1}}, \tensor*[^{1}]{a}{_{2}}, \tensor*[^{1}]{a}{_{3}})} g( \tensor*[^{1}]{a}{_{1}}, \tensor*[^{1}]{a}{_{2}}, \tensor*[^{1}]{a}{_{3}}) = 0$.
Solving this equation we obtain,
\begin{equation}
\begin{split}
	\left[
	\begin{array}{c}
		\frac{1}{3}\; \al_1 \Big(\frac{1}{3} \big(\tensor*[^{1}]{a}{_{1}} + \tensor*[^{2}]{a}{_{1}^*} +
		   \tensor*[^{3}]{a}{_{1}^*}\big)\Big)^{\al_1-1}
		   - \frac{1}{3} \big( \tensor*[^{2}]{\pi}{_{1}^*} - \tensor*[^{3}]{\pi}{_{1}^*} \big) \\
		   - \frac{1}{3}\;\Be_2 \Big(\frac{1}{3} \big(\tensor*[^{1}]{a}{_{2}} + \tensor*[^{2}]{a}{_{2}^*} +
		   \tensor*[^{3}]{a}{_{2}^*}\big)\Big)^{\Be_2-1} 
		   - \frac{1}{3}  \big( \tensor*[^{2}]{\pi}{_{2}^*} - \tensor*[^{3}]{\pi}{_{2}^*} \big) \\
		   -\frac{1}{3}\;\Be_3 \Big(\frac{1}{3} \big(\tensor*[^{1}]{a}{_{3}} + \tensor*[^{2}]{a}{_{3}^*} +
		   \tensor*[^{3}]{a}{_{3}^*}\big)\Big)^{\Be_3-1} 
		   - \frac{1}{3} \big( \tensor*[^{2}]{\pi}{_{3}^*} - \tensor*[^{3}]{\pi}{_{3}^*} \big)
	\end{array}
	\right]
	= 0. 
	\label{grad_g_a_pi}
	\end{split}
\end{equation} 
Substituting the values of $\tensor*[^{i}]{a}{_{j}^*}, \tensor*[^{i}]{\pi}{_{j}^*}, i,j\in\{1,2,3\},$ from (\ref{msg_prof}) into (\ref{grad_g_a_pi}) implies that,
\begin{equation}
\begin{split}
	\left[
	\begin{array}{c}
		\al_1 \Big(\frac{1}{3} \big(\tensor*[^{1}]{a}{_{1}} + 2 \h{a}_1^{*} \big)\Big)^{\al_1-1}
		   - 2 \Be_1 \h{a}_1^{*^{(\Be_1-1)}} \\
		   - \Be_2 \Big(\frac{1}{3} \big(\tensor*[^{1}]{a}{_{2}} + 2 \h{a}_2^{*} \big)\Big)^{\Be_2-1} 
		   + \Be_2 \h{a}_2^{*^{(\Be_2-1)}} \\
		   -\Be_3 \Big(\frac{1}{3} \big(\tensor*[^{1}]{a}{_{3}} + 2 \h{a}_3^{*} \big)\Big)^{\Be_3-1} 
		   + \Be_3 \h{a}_3^{*^{(\Be_3-1)}}
	\end{array}
	\right]
	= 0. 
	\label{grad_g_a*}
	\end{split}
\end{equation} 
Substituting the values of $\h{a}_1^{*},\h{a}_2^{*},\h{a}_3^{*}$ from (\ref{a*_cent}) into (\ref{grad_g_a*}), and solving for $\tensor*[^{1}]{a}{_{1}}, \tensor*[^{1}]{a}{_{2}}, \tensor*[^{1}]{a}{_{3}}$, we obtain, 
\begin{equation}
\begin{split}
	\left[
	\begin{array}{c}
		\tensor*[^{1}]{a}{_{1}} \\
		\tensor*[^{1}]{a}{_{2}} \\
		\tensor*[^{1}]{a}{_{3}} 
	\end{array}
	\right]
	= 
	\left[
	\begin{array}{c}
		3 \left( \frac{2\Be_1}{\al_1} \left( \frac{\al_1}{2\Be_1} \right)^{\frac{\Be_1-1}{\Be_1-\al_1}} \right)^{\frac{1}{\al_1-1}}
		- 2 \left( \frac{\al_1}{2\Be_1} \right)^{\frac{1}{\Be_1-\al_1}} \\
		3 \left( \h{a}_2^{*^{(\Be_2-1)}} \right)^{\frac{1}{\Be_2-1}} 
		- 2 \h{a}_2^* \\
		3 \left( \h{a}_3^{*^{(\Be_3-1)}} \right)^{\frac{1}{\Be_3-1}} 
		- 2 \h{a}_3^*
	\end{array}
	\right]
	=
	\left[
	\begin{array}{c}
			\left( \frac{\al_1}{2\Be_1} \right)^{\frac{1}{\Be_1-\al_1}} \\
			\h{a}_2^* \\
			\h{a}_3^*
	\end{array}
	\right]
	=
		\left[
	\begin{array}{c}
		\tensor*[^{1}]{a}{_{1}^*} \\
		\tensor*[^{1}]{a}{_{2}^*} \\
		\tensor*[^{1}]{a}{_{3}^*}
	\end{array}
	\right]
	\label{grad_g_sol}
	\end{split}
\end{equation} 
The last equality in (\ref{grad_g_sol}) follows from the definition of $\tensor*[^{1}]{a}{_{1}^*}, \tensor*[^{1}]{a}{_{2}^*}, \tensor*[^{1}]{a}{_{3}^*}$ in (\ref{msg_prof}) and the value of the centralized optimum in (\ref{a*_cent}). Note that for $(\tensor*[^{1}]{a}{_{1}}, \tensor*[^{1}]{a}{_{2}}, \tensor*[^{1}]{a}{_{3}}) = (\tensor*[^{1}]{a}{_{1}^*}, \tensor*[^{1}]{a}{_{2}^*}, \tensor*[^{1}]{a}{_{3}^*})$, the last three quadratic terms in (\ref{u1_t1_expanded}) are zero for all $(\tensor*[^{1}]{\pi}{_{1}}, \tensor*[^{1}]{\pi}{_{2}}, \tensor*[^{1}]{\pi}{_{3}})$ (since $\tensor*[^{1}]{a}{_{1}^*}=\tensor*[^{2}]{a}{_{1}^*}, \ \tensor*[^{1}]{a}{_{2}^*}=\tensor*[^{2}]{a}{_{2}^*}, \ \tensor*[^{1}]{a}{_{3}^*}=\tensor*[^{2}]{a}{_{3}^*}$ by (\ref{msg_prof})). Hence, any price proposal  $(\tensor*[^{1}]{\pi}{_{1}}, \tensor*[^{1}]{\pi}{_{2}}, \tensor*[^{1}]{\pi}{_{3}})$ along with   $(\tensor*[^{1}]{a}{_{1}^*}, \tensor*[^{1}]{a}{_{2}^*}, \tensor*[^{1}]{a}{_{3}^*})$ is a maximizer of (\ref{u1_t1_expanded}). Consequently, the message $\Bm_1^*$ defined in (\ref{msg_prof}) is a maximizer in (\ref{NE_argmax_cond_1}).

By following similar arguments as above, the conditions (\ref{NE_argmax_cond_2}) and (\ref{NE_argmax_cond_3}) can also be derived. Thus, the message profile $\Bm^* = (\Bm_1^*, \Bm_2^*, \Bm_3^*)$ is a NE of the game under consideration. 
\\

We now show that the allocations obtained at the above NE possess the properties established by Theorem~\ref{chap5_Thm_NE_opt}.

By substituting the messages defined by (\ref{msg_prof}) into equation (\ref{chap5_Eq_a_hat_i})
%\textcolor{red}{check} 
we obtain,
\begin{equation}
	\left[
	\begin{array}{c}
			\h{a}_1(\Bm^*) \\
			\h{a}_2(\Bm^*) \\
			\h{a}_3(\Bm^*)		
	\end{array}
	\right]
	=
	\left[
	\begin{array}{c}
			\frac{1}{3} \big(\tensor*[^{1}]{a}{_{1}^*} + \tensor*[^{2}]{a}{_{1}^*} + \tensor*[^{3}]{a}{_{1}^*}\big) \\
			\frac{1}{3} \big(\tensor*[^{1}]{a}{_{2}^*} + \tensor*[^{2}]{a}{_{2}^*} + \tensor*[^{3}]{a}{_{2}^*}\big) \\
			\frac{1}{3} \big(\tensor*[^{1}]{a}{_{3}^*} + \tensor*[^{2}]{a}{_{3}^*} + \tensor*[^{3}]{a}{_{3}^*}\big)		
	\end{array}
	\right]	
	=
		\left[
	\begin{array}{c}
			\h{a}_1^* \\
			\h{a}_2^* \\
			\h{a}_3^*		
	\end{array}
	\right].
	\label{NE_alloc_cent}
\end{equation}
Thus (\ref{NE_alloc_cent}) shows that the allocation at Nash equilibrium $\Bm^*$ is the optimal centralized allocation $\h{\ba}^*$. 
\\

To see budget balance at $\Bm^*$, note that the last six terms on the RHS of (\ref{tax1_expanded}) are zero at $\Bm=\Bm^*$. Similarly, the corresponding terms in the taxes of users 2 and 3 will also be zero. Adding the remaining terms in the users' taxes gives,
\begin{equation}
\begin{split}
	\textstyle{\sum_{i\in\{1,2,3\}}} \h{t}_i(\Bm^*) = 
	&	\ \ \ \, \big( \tensor*[^{2}]{\pi}{_{1}^*} - \tensor*[^{3}]{\pi}{_{1}^*} \big) \; \h{a}_1^*
	+ \big( \tensor*[^{2}]{\pi}{_{2}^*} - \tensor*[^{3}]{\pi}{_{2}^*} \big) \; \h{a}_2^*
	+ \big( \tensor*[^{2}]{\pi}{_{3}^*} - \tensor*[^{3}]{\pi}{_{3}^*} \big) \; \h{a}_3^* \\
	&
	+	\big( \tensor*[^{3}]{\pi}{_{1}^*} - \tensor*[^{1}]{\pi}{_{1}^*} \big) \; \h{a}_1^*
	+ \big( \tensor*[^{3}]{\pi}{_{2}^*} - \tensor*[^{1}]{\pi}{_{2}^*} \big) \; \h{a}_2^*
	+ \big( \tensor*[^{3}]{\pi}{_{3}^*} - \tensor*[^{1}]{\pi}{_{3}^*} \big) \; \h{a}_3^* \\
	&
	+	\big( \tensor*[^{1}]{\pi}{_{1}^*} - \tensor*[^{2}]{\pi}{_{1}^*} \big) \; \h{a}_1^*
	+ \big( \tensor*[^{1}]{\pi}{_{2}^*} - \tensor*[^{2}]{\pi}{_{2}^*} \big) \; \h{a}_2^*
	+ \big( \tensor*[^{1}]{\pi}{_{3}^*} - \tensor*[^{2}]{\pi}{_{3}^*} \big) \; \h{a}_3^* \\
	= & \ \ \ \, 0.	
	\end{split}
\end{equation}

%To see voluntary participation at $\Bm^*$, we compute each user's aggregate utility at $\Bm^*$.
%For user 1 it gives,
%\begin{equation}
%\begin{split}
%	& u_1(\h{\ba}(\Bm^*)) - \h{t}_1(\Bm^*) =  \\
%	& {\h{a}_1^*}^{\al_1} - {\h{a}_2^*}^{\Be_2} - {\h{a}_3^*}^{\Be_3}		
%\end{split}
%\end{equation}
%\textcolor{red}{should we provide this derivation?}
Voluntary participation at $\Bm^*$ can be seen by substituting $\Bm^*$ from (\ref{msg_prof}) into $u_i(\h{\ba}(\Bm^*)) - \h{t}_i(\Bm^*), \; i\in\{1,2,3\}$, and computing this aggregate utility for each user using (\ref{u_i_def}), (\ref{a*_cent}), (\ref{msg_prof}) and (\ref{chap5_Eq_a_hat_i})--(\ref{chap5_Eq_l_ij}). The computation shows that $u_i(\h{\ba}(\Bm^*)) - \h{t}_i(\Bm^*) \geq 0 \fa i\in\{1,2,3\}$ which means that the users obtain at least zero utility at $\Bm^*$ which is same as the utility they would obtain if they do not participate in the mechanism. 
\\

%%%%%%%%%%%%%%%%%%%%%%%%%%%%%%%%%%%%%%%%%%%%%%%%%%%%%%%%%%%%%%%%%%%%%%%%%%%%%%%%%%%%%%%%%%%
%%%%%%%%%%%%%%%%%%%%%%%%%%%%%%%%%%%%%%%%%%%%%%%%%%%%%%%%%%%%%%%%%%%%%%%%%%%%%%%%%%%%%%%%%%%
%%%%%%%%%%%%%%%%%%%%%%%%%%%%%%%%%%%%%%%%%%%%%%%%%%%%%%%%%%%%%%%%%%%%%%%%%%%%%%%%%%%%%%%%%%%

\end{document}